\documentclass[pre,twocolumn,showpacs,preprintnumbers]{revtex4}
\usepackage{bm}
\usepackage{amssymb}
\usepackage{amsmath}
\usepackage{graphicx}

\begin{document}
\title{Direct MD simulation of liquid-solid phase equilibria for two-component plasmas}

\author{A. S. Schneider}\email{andschn@umail.iu.edu}
\author{J. Hughto}\email{jhughto@astro.indiana.edu}
\author{C. J. Horowitz}\email{horowit@indiana.edu} 
\affiliation{Department of Physics and Nuclear Theory Center,
             Indiana University, Bloomington, IN 47405}
\author{D. K. Berry}
\affiliation{University Information Technology Services,
             Indiana University, Bloomington, IN 47408}

\date{\today}
\begin{abstract}
We determine the liquid-solid phase diagram for carbon-oxygen and oxygen-selenium plasma mixtures using two-phase MD simulations.  We identified liquid, solid, and interface regions using a bond angle metric. To study finite size effects, we perform 27648 and 55296 ion simulations.  To help monitor non-equilibrium effects, we calculate diffusion constants $D_i$.  For the carbon-oxygen system we find that $D_O$ for oxygen ions in the solid is much smaller than $D_C$ for carbon ions and that both diffusion constants are 80 or more times smaller than diffusion constants in the liquid phase.  There is excellent agreement between our carbon-oxygen phase diagram and that predicted by Medin and Cumming.  This suggests that errors from finite size and non-equilibrium effects are small and that the carbon-oxygen phase diagram is now accurately known.   The oxygen-selenium system is a simple two-component model for more complex rapid proton capture nucleosynthesis ash compositions for an accreting neutron star.  Diffusion of oxygen, in a predominately selenium crystal, is remarkably fast, comparable to diffusion in the liquid phase.  We find a somewhat lower melting temperature for the oxygen-selenium system than that predicted by Medin and Cumming.  This is probably because of electron screening effects.

\end{abstract}
\smallskip
\pacs{97.20.Rp  
, 64.70.D- 
, 64.70.dg 
}
\maketitle

\section{Introduction}
\label{Introduction}

Observations of cooling White Dwarf (WD) stars provide important information on the ages and evolution of stellar systems \cite{cosmochron, garcia-berro, renedo, salaris1}.  The interior of a WD is a Coulomb plasma of ions and a degenerate electron gas.  As the star cools this plasma crystallizes.  This can delay WD cooling, see for example ref. \cite{salaris}.  Winget et al. recently observed effects from the latent heat of crystallization on the luminosity function of WDs in the globular cluster NGC 6397 \cite{winget}.   Winget et al.'s observations may constrain the melting temperature of the carbon and oxygen mixtures expected in these WD cores.  In addition, astroseismology provides an alternative way to study crystallization in WD, see for example \cite{astroseismology}.

Furthermore, material accreting onto a neutron star (NS) will freeze to form new NS crust.  A variety of nuclear reactions can take place, including rapid proton capture nucleosynthesis (rp process \cite{rpash,rpash2}) followed by electron captures, as the material is advected to higher densities.  Horowitz {\it et al.} studied the crystallization of a complex rp process ash consisting of 17 chemical elements from oxygen to selenium \cite{HBB}.  They found chemical separation upon freezing, with low $Z$ elements preferentially remaining in the liquid NS ocean while high $Z$ elements crystallize to form new NS crust.  This change in composition of the ocean may be important for superbursts.  These are thought to be energetic thermonuclear explosions of carbon \cite{superbursts, superbursts2, superbursts3}.

Melting temperatures and other properties of the liquid-solid phase diagram for multi-component plasmas have been determined from computer simulations.  Segretain et al. calculated the carbon-oxygen phase diagram assuming a local density model for the free energy of the solid \cite{segretain}.  While, Ogata et al. \cite{ogata},\cite{ichimaru} and DeWitt et al. \cite{dewitt03},\cite{dewitt96} calculated the phase diagram based on Monte Carlo or Molecular Dynamics (MD) simulation free energies for both the liquid and solid phases.  Recently Potekhin et al. have made accurate calculations of the free energy of liquid mixtures \cite{potekhin09},\cite{potekhin09b} and Medin and Cumming calculated the phase diagram for both binary mixtures such as C/O and much more complicated multi-component mixtures \cite{medin}.

All of these works determine liquid-solid phase equilibria by equating liquid and solid free energies that have been calculated separately.  This procedure allows the use of smaller Monte Carlo or MD simulations where only a single phase is present at a time.  However, it may be very sensitive to any small errors in the free energy difference between liquid and solid phases.  Indeed for the C/O system, Segretain et al. predict higher melting temperatures and a spindle type phase diagram while both Ogata et al. and Medin and Cumming predict lower melting temperatures and an azeotrope type phase diagram.  In a spindle-type phase diagram the melting temperature of the mixture is always greater than the melting temperature of pure carbon, while in an azeotrope type phase diagram the melting temperature of the mixture can be lower than that of pure carbon.  This difference in phase diagrams could be due to small errors in Segretain's solid free energies. 

Furthermore, equating the free energies of liquid and solid phases provides no information on the dynamics of the phase transition.  For example, although there have been some studies of nucleation for one component plasma systems \cite{nucleation1,nucleation2}, there have been almost no studies of nucleation for multi-component plasmas.  Finally, interface properties are not addressed.  For one component systems there has been some work on surface properties, see for example \cite{surface, surface2}.   For multi-component systems there is, in general, a gradient in composition across the liquid-solid interface.  However, the spatial extent of this gradient has not been determined.       

Recently, we performed direct two phase molecular dynamics simulations of liquid-solid phase equilibria both for carbon / oxygen mixtures in WD \cite{WDPRL} and for a complex 17 component mixture modeling the crust of an accreting neutron star \cite{HBB}.  These simulations have both liquid and solid phases present simultaneously.  This allows a direct determination of the melting temperature, and the composition of the liquid and solid phases from a single simulation.  Furthermore, phase equilibria for very complicated systems can be simulated in this way.  

However, direct simulations need to address potential systematic errors from finite size effects and from the lack of thermodynamic equilibration.  Finite size effects are potentially important because one must fit not only liquid and solid phases but also two liquid-solid interfaces within the simulation volume.  This, in general, requires a larger simulation volume than for simulations of only a single phase.  However, recent computer advances have dramatically reduced the computational limitations on these larger simulations.  It is now ``easy" to simulate much larger systems than have typically been run in the past.

One must run these two-phase simulations long enough to insure that the phases have come into thermodynamic equilibrium.  This may require impurities to diffuse throughout the solid phase.  However, diffusion in the solid phase is relatively fast, for these Coulomb systems, because the ions do not have hard core interactions.  There is only a relatively soft $1/r$ interaction between ions.  As a result, ions can move past one another.  We have studied diffusion in Coulomb crystals in a recent paper \cite{soliddiffusion}.  If finite size and non-equilibrium effects are addressed, direct two phase simulations should yield accurate results.  The systematic errors from two-phase simulations are likely very different from previous free energy calculations.  Therefore, comparing the two methods provides an important check on both approaches.

In the laboratory, one can observe complex (or dusty) plasma crystals.  Complex plasmas (CP) are low temperature plasmas containing charged micro-particles, for a review see Fortov et al. \cite{fortov}.  Often the micro-particles are micron sized spheres that acquire large electric charges and the strong Coulomb interactions between micro-particles can lead to crystallization.  Indeed plasma crystals were first observed in the laboratory in 1994 \cite{dusty_plasma}.   Alternatively, one can study systems of charged colloidal spheres.  For example, Lorenz and Palberg observed melting and freezing lines for a binary mixture of colloidal spheres \cite{Lorenz_Palberg}.

In this paper we study freezing of binary mixtures of both C/O and O/Se.  The C/O system is important for White Dwarfs while the O/Se system provides a simple binary model of the complex rp ash composition in accreting NS.   We perform MD simulations with both 27648 and 55296 ions.  This allows us to study finite size effects.  We discuss our MD formalism in Section \ref{Formalism},  present results in Section \ref{CO} for the carbon-oxygen, and Section \ref{results_oxygen_selenium} for the oxygen-selenium system, and conclude in Section \ref{Conclusions}. 

\section{Formalism}
\label{Formalism}

In Section \ref{subsec.MD} we describe our two-phase MD simulation formalism.  This is very similar to what we used earlier for the freezing of rapid proton capture nucleosynthesis ash on accreting neutron stars \cite{HBB} and for carbon and oxygen mixtures in WD \cite{WDPRL}.  Next in Section \ref{subsec.phase} we describe our analysis procedure for determining if a given region of the simulation is in a liquid or solid phase.
   
\subsection{MD formalism}
\label{subsec.MD}

We consider a system of ions, of two different charges, and electrons.  The electrons are assumed to form a degenerate Fermi gas.  The ions are fully pressure ionized and interact with each other via screened Coulomb interactions.  The potential between the $i$th and $j$th ion is assumed to be,
\begin{equation}
v_{ij}(r)=\frac{Z_iZ_j e^2}{r} {\rm e}^{-r/\lambda}.
\label{v(r)}
\end{equation}
Here the ion charges are $Z_i$ and $Z_j$, $r$ is their separation and the electron screening length is $\lambda$.  For cold relativistic electrons, the Thomas Fermi screening length is $\lambda^{-1}=2\alpha^{1/2}k_F/\pi^{1/2}$ where the electron Fermi momentum $k_F$ is $k_F=(3\pi^2n_e)^{1/3}$ and $\alpha$ is the fine structure constant.  Finally the electron density $n_e$ is equal to the ion charge density, $n_e=\langle Z\rangle n$, where $n$ is the ion density and $\langle Z\rangle$ is the average charge.  We cutoff the potential for $r>8\lambda$.  Our simulations are classical and we have neglected the electron mass (extreme relativistic limit).   This is to be consistent with our previous work on neutron stars.  However, the electron mass is important at the lower densities in WD and this may change our results slightly \cite{pot1}.  Also quantum effects could play some role at high densities \cite{pot2},\cite{jones}.

The simulations can be characterized by an average Coulomb parameter $\Gamma$,
\begin{equation}
\Gamma= \frac{\langle Z^{5/3} \rangle e^2}{a_e T}\, .
\label{gammamix}
\end{equation} 
Here $\langle Z^{5/3} \rangle$ is an average over the ion charges, $T$ is the temperature, and the electron sphere radius $a_e$ is $a_e=(3/4\pi n_e)^{1/3}$.

Time can be measured in units of one over the plasma frequency $\omega_p$.  Long wavelength fluctuations in the charge density can undergo oscillations at the plasma frequency.  This depends on the ion charge $Z$ and mass $M$.  For mixtures we define a hydrodynamical plasma frequency $\bar\omega_p$ from the simple averages of $Z$ and $M$,
\begin{equation}
\bar\omega_p=\Bigl[\frac{4\pi e^2\langle Z\rangle^2 n}{\langle M \rangle}\Bigr]^{1/2}.
\label{omega}
\end{equation}
Note that other choices for the average over composition in Eq. \ref{omega} are possible.  However, they are expected to give very similar results for the average plasma frequency.

All of our carbon-oxygen simulations are run for the same electron density of $n_e=5.026\times 10^{-4}$ fm$^{-3}$, while the oxygen-selenium simulations are run for $n_e=2.254\times 10^{-3}$ fm$^{-3}$. Since the pressure is dominated by the electronic contribution, constant electron density corresponds, approximately, to constant pressure.  The density can be scaled to any desired value by also changing the temperature $T$ so that the value of $\Gamma$, see Eq. \ref{gammamix}, remains the same.    

\subsection{Interface finding algorithm}
\label{subsec.phase}

In this Section we describe an algorithm for specifying if a given region of the simulation belongs to the bulk liquid, or bulk solid, phase or if the region belongs to a liquid-solid interface.   Often determining whether a cluster of ions is a liquid or a bcc solid is simple when visually inspected.  However this determination is difficult to obtain numerically.  For an entire system, phase determination can be accomplished by computing the global order parameter $Q_6$ \cite{steinhardt83}, see below. In this work, we use the prescription laid out by ref. \cite{tenwolde96} to determine whether individual ions are liquid-like or solid-like.

For each ion $i$, an ion $j$ is defined as neighbors if it is within a given radius $r_{cut}=4a$ with $a$ the ion sphere radius $a=(3/4\pi n)^{1/3}$.  The vectors $\hat{\textbf{r}}_{ij}$ joining neighbors are called bonds.  The direction of these vectors can be described by $\theta_{ij}$ and $\phi_{ij}$ in the frame of ion $i$.  The local structure around ion $i$ can be characterized using spherical harmonics $Y_{lm}(\theta_{ij},\phi_{ij})$ by 
\begin{equation}
\bar{q}_{lm}(i) = \frac{\sum_{j=1}^{N_b(i)}  \alpha(r_{ij})Y_{lm}(\theta_{ij},\phi_{ij})}{\sum_{i=1}^{N_b(i)}  \alpha(r_{ij})}
\end{equation}
where $N_b(i)$ is the number of ions bonded with ion $i$ and
\begin{equation}
\alpha(r_{ij}) = \left(\frac{r_{ij}-4a}{4a}\right)^2
\end{equation}
if $r_{ij}<4a$ and $\alpha(r_{ij})=0$ otherwise. 

These local order parameters are large in both the solid and the liquid.  The global order parameter $Q_6$ is calculated from an average over all of the $N$ ions,
\begin{equation}
q_{6m}=\frac{1}{N}\sum_{i=1}^N\bar q_{6m}(i),
\end{equation}
\begin{equation}
Q_6=\Bigl[\frac{4\pi}{13}\sum_{m=-6}^6 q_{6m}^*q_{6m}\Bigr]^{1/2}.
\end{equation}
This is large in the solid due to the fact that the $\bar{q}_{6m}(i)$ add up coherently.  In the liquid, $\bar{q}_{6m}(i)$ add incoherently, so $Q_6$ is small.  This coherence is exploited to determine local order.

For each $\bar{q}_{6m}(i)$ a normalization is applied,
\begin{equation}
\tilde{q}_{6m}(i)\equiv\frac{\bar{q}_{6m}(i)}{\displaystyle\left[\sum_{m=-6}^6\left|\bar{q}_{6m}(i)\right|^2\right]^{1/2}}
\end{equation}
A dot product can now be defined of the vectors $\textbf{q}_6$ for neighboring particles $i$ and $j$,
\begin{equation}
\textbf{q}_6(i)\cdot\textbf{q}_6(j)\equiv\sum_{m=-6}^6\tilde{q}_{6m}(i)\tilde{q}_{6m}^*(j).
\end{equation}
By construction, $\textbf{q}_6(i)\cdot\textbf{q}_6(i)=1$.

We use the same criterion as ref. \cite{tenwolde96} for determining whether two particles are connected, namely $\textbf{q}_6(i)\cdot\textbf{q}_6(j)>0.5$.  This criterion will be met for most of the bonds in the solid.  In the liquid, two neighbors may be in phase and considered connected, but that is unlikely to be true for all of the neighbors.  Therefore, we use a threshold on the number of connections to determine if an ion is solid-like or liquid-like.  This threshold is 20 connections.  An ion is considered solid-like if it is connected to 20 or more neighbors.  Otherwise the ion is considered liquid-like.  Note that on average an ion in the carbon-oxygen system has about 62 neighbors.

Now that each ion is tagged as either solid-like or liquid-like, the interface in our two-phase simulations can be found.  Deep in the solid, a vast majority of the ions within a certain radius of a given ion are identified as solid-like.  In the bulk of the liquid a similar majority is identified as liquid-like.  Along the interface, there is a mixture of solid-like and liquid-like ions.  For this reason, we tag an ion as being in the solid or liquid if a large majority ($> 80\%$) of the ions within $4a$ of a given ion are the same phase.  If this criterion is not met, then the ion is determined to be in the interface.  Performing this identification leads to results that will be shown in Fig. \ref{Fig2} of Section \ref{CO}.  Notice that ions determined to be in the interface are found where one would expect them, along the border of the solid and liquid phases.

Note that this procedure is slightly modified for the oxygen-selenium system. Bonds involving oxygen ions, in a predominately selenium crystal, are not as well ordered as selenium-selenium bonds.  Therefore, for this system we only consider selenium-selenium bonds, as discussed in Sec. \ref{results_oxygen_selenium}.  

\section{Results for carbon and oxygen system}
\label{CO}
In this section we present results for the phase diagram and diffusion constants for carbon and oxygen systems.  Our previous carbon and oxygen results in ref. \cite{WDPRL} were based on MD simulations with 27648 ions.  Here we perform three larger simulations with 55296 ions in order to study finite size effects.  In addition we calculate diffusion constants in order to monitor non-equilibrium effects. 

To minimize finite size effects, we use a rectangular simulation volume that is twice as long in the $z$ direction compared to the $x$ or $y$ directions.   We use periodic boundary conditions in this rectangular box.  Note that, we evaluate the interaction between two particles as the single interaction with the nearest periodic image.  We do not include an Ewald sum over further periodic images because our box is so large that interactions with periodic images other than the nearest one are very small.  The initial conditions consist of a cube of crystalline phase that is stacked together in the $z$ direction with an equal sized cube of liquid phase.  This rectangular geometry increases the distance between the two liquid-solid interfaces compared to a cubical simulation volume. 

\begin{figure}[ht]
\begin{center}
\includegraphics[width=6.5in,angle=90,clip=true] {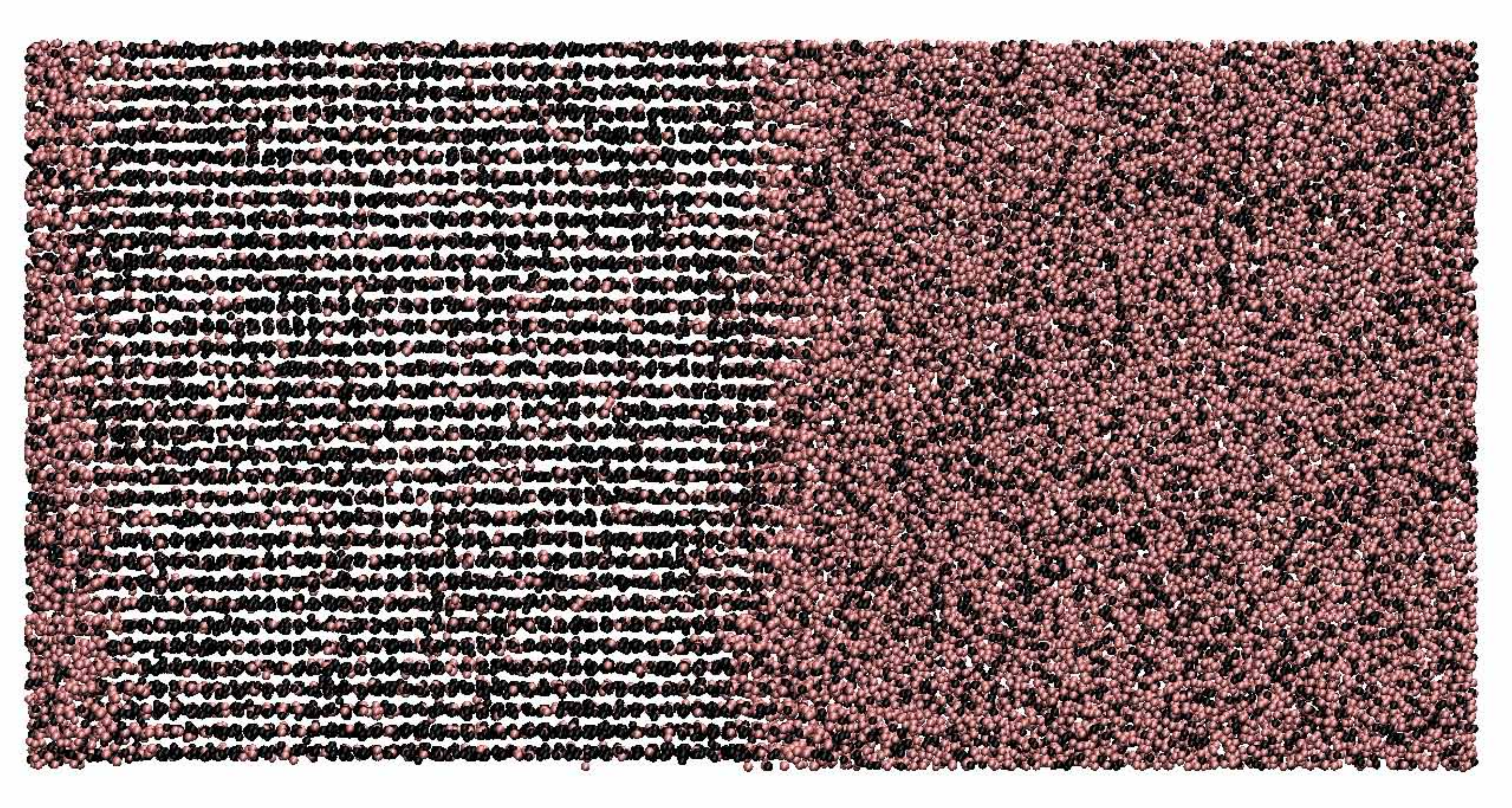}
\caption{(Color on line) Final configuration of carbon ions (light red) and oxygen ions (black) in a 55296 ion simulation that consisted of 75\% oxygen and 25\% carbon.}
\label{Fig1}
\end{center}
\end{figure}

\subsection{Run with 75\% oxygen}
\label{subsec.O75}

Our first simulation has an average composition of 75\% oxygen and 25\% carbon (by number).  We independently prepare solid and liquid initial conditions and then combine them to obtain the full 55296 ion initial conditions.  Based on our earlier results suggesting that the solid phase will be enriched in oxygen from \cite{WDPRL}, we prepare the liquid configuration with 70\% oxygen, 30\% carbon and the solid configuration with 80\% oxygen, 20\% carbon.   We start with a 3456 ion configuration with random positions and velocities and cool the system until it solidifies.  We then combine 8 copies of this 3456 ion solid to make a 27648 ion solid configuration and continue to evolve this 27648 ion system for a time $t\bar\omega_p=8900$ at $\Gamma=213.1$.  We form a liquid configuration by starting with an independent 3456 ion system with random coordinates and evolve the system for a time $t\bar\omega_p=4400$ at the same $\Gamma$.  We combine 8 copies of this liquid configuration to make a 27648 ion liquid and evolve for a further $t\bar\omega_p=8800$. Finally we combine the 27648 ion solid configuration with the 27648 ion liquid configuration to form the full 55296 ion initial condition.

\begin{figure}[ht]
\begin{center}
\includegraphics[width=6.5in,angle=90,clip=true] {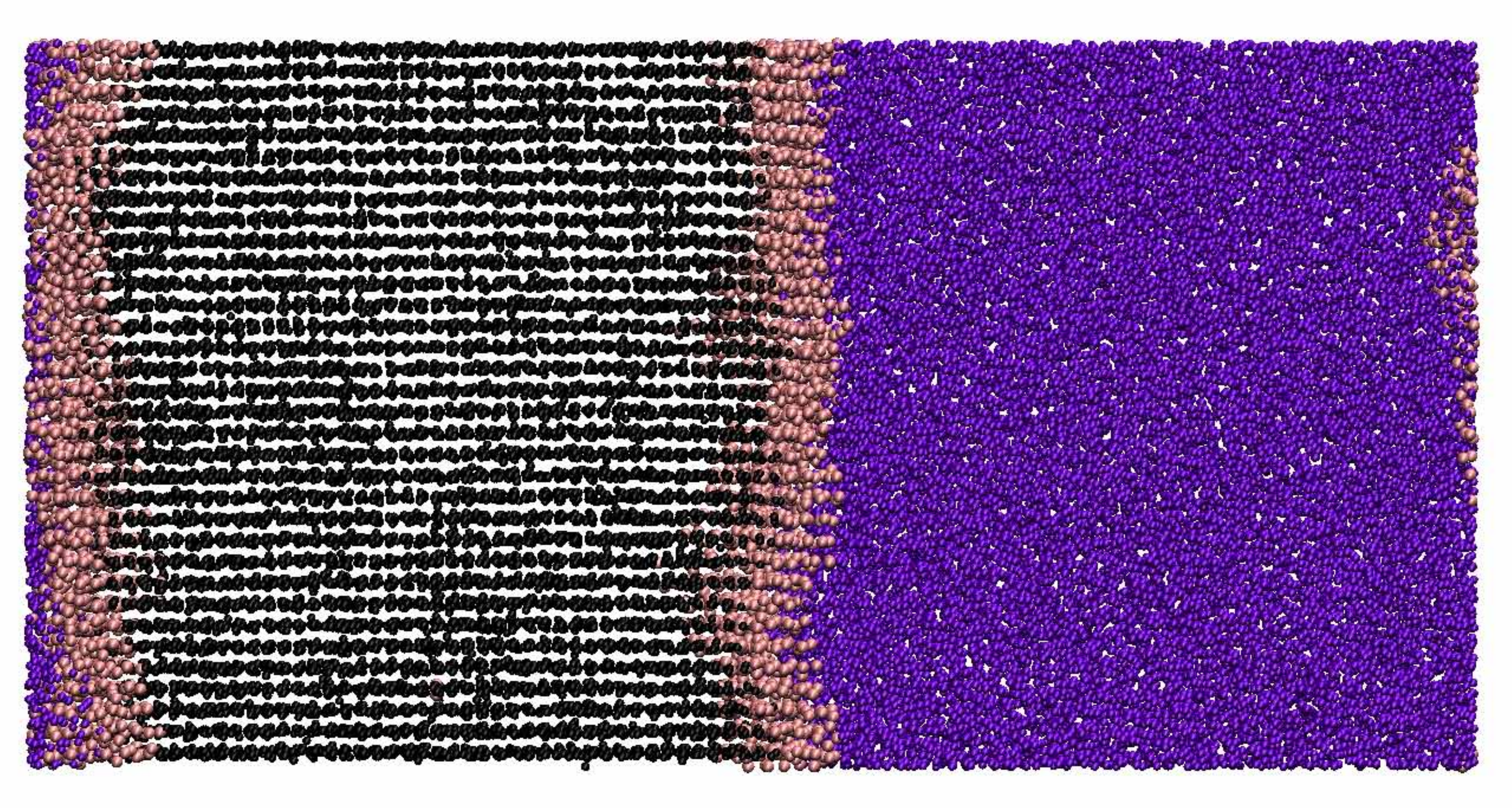}
\caption{(Color on line) Final configuration of liquid (dark purple), interface (light red), and solid (black in bottom half) ions from a 55296 ion simulation that consisted of 75\% oxygen and 25\% carbon.  These liquid, solid and interface, regions are determined using a bond angle metric as discussed in Sec. \ref{subsec.phase}.}
\label{Fig2}
\end{center}
\end{figure}
   
This initial condition is not equilibrated for a number of reasons.  First the two interfaces between liquid and solid may have high energies because we have simply combined two different initial conditions.  There may be liquid and solid ions close to each other.  In addition, the solid part of the initial condition may not be equilibrated because it has carbon and oxygen ions positioned on lattice sites more or less at random.  The equilibrated solid may have important correlations between ions of different charges.  Finally, the compositions of the liquid and solid phases may be wrong.   Therefore, carbon ions may diffuse into or out of the solid region until the composition of the liquid and solid phases equilibrate.  This may require evolving the system for a considerable time.

We evolved the full 55296 ion system for a total time $2.8\times 10^6/\bar\omega_p$ using a velocity Verlet \cite{verlet} time step of $0.177/\bar\omega_p$.  We employed a simple hybrid Open MP / MPI computer code and use about 768 cores on the Cray XT5 system Kraken \cite{Kraken}.  This evolution took a total of about 2 weeks. During the run we rescaled the velocities every 100 time steps in order to keep the temperature approximately constant.  We slowly adjust the temperature so that approximately half of the system remains liquid and half solid.  In Fig. \ref{Fig1} we show the final configuration of the 55296 ions.  The bottom half of the simulation volume is seen to contain a crystalline region.  We use the procedure of Sec. \ref{subsec.phase} to determine which parts of the system are liquid, solid, or belong to the two liquid-solid interfaces.  This is shown in Fig. \ref{Fig2}.  Note that the interface regions are not rectangular and show some fluctuations.

\begin{figure}[ht]
\begin{center}
\includegraphics[width=3.5in,angle=0,clip=true] {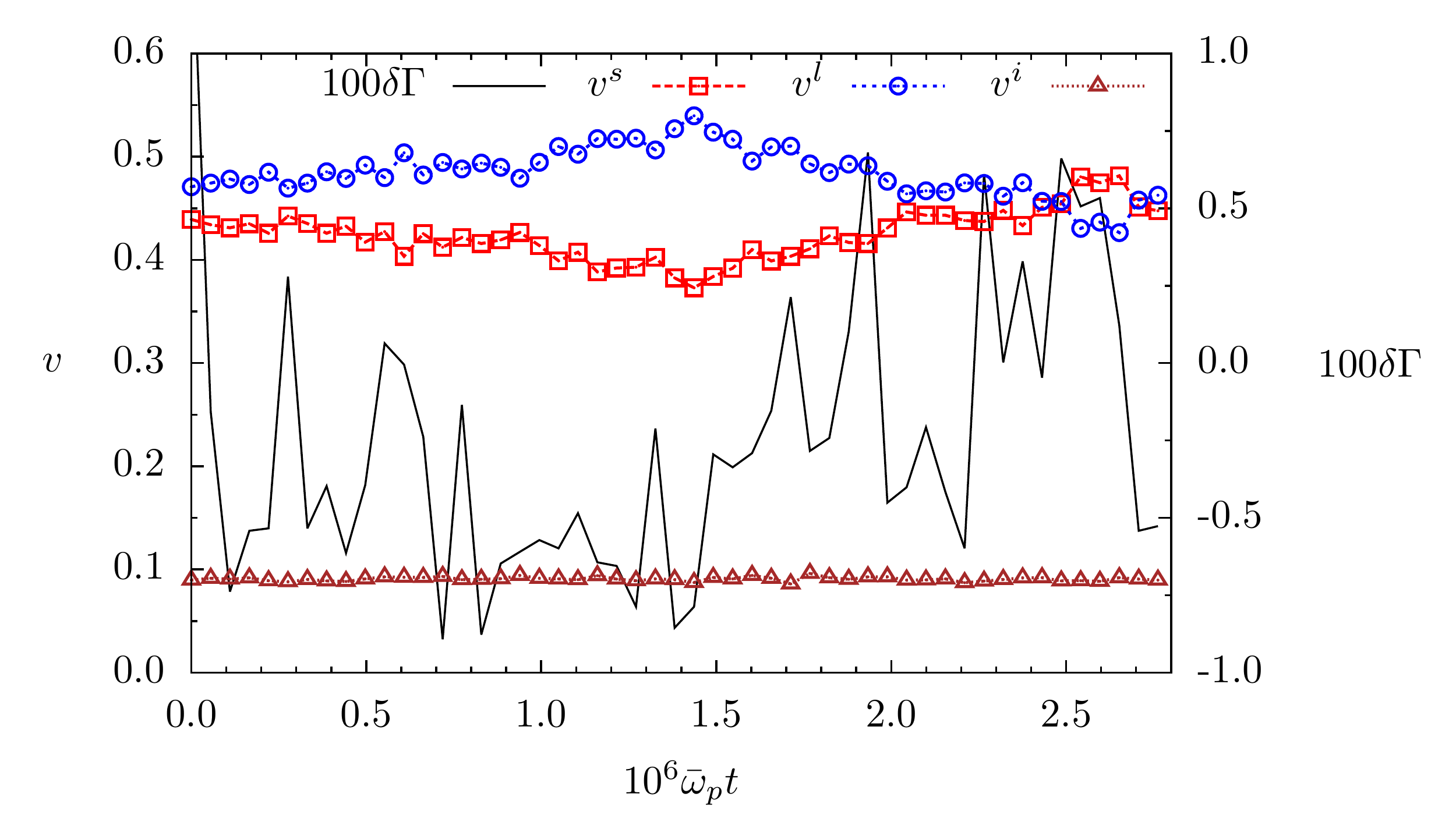}
\caption{(Color on line) Fractional fluctuations in Coulomb parameter $\delta\Gamma$ (solid black line and right hand scale), see Eq. \ref{deltaGamma} vs simulation time $t$ for a 55296 ion system with 75\% oxygen and 25\% carbon.  Also shown are the fraction of the system that is solid  (red squares), liquid (blue circles) or interface (brown triangles).}
\label{Fig3}
\end{center}
\end{figure}

In Fig. \ref{Fig3} we show the fluctuations in $\Gamma$ (or equivalently one over the temperature) during the run.  We calculate the final $\bar\Gamma$ value as the average of $\Gamma$ over the last third of the run, see Table \ref{tableone}.  We define the scaled fluctuations  in $\Gamma$ as $\delta\Gamma$,
\begin{equation}
\delta \Gamma=\frac{\Gamma-\bar\Gamma}{\bar\Gamma}.
\label{deltaGamma}
\end{equation}
As the run started, it was necessary to use a low temperature (high $\Gamma$) in order to keep the badly non-equilibrated solid frozen.  However as the system rapidly equilibrated the temperature could be quickly reduced towards its final equilibrated value.  The temperature is then seen to fluctuate around this value at the half percent level or less.  Fig. \ref{Fig3} also shows the fraction of the system that is solid, liquid, or interface vs time.  The interface fraction is very constant (since the thickness of the interface depends on our definition in Sec. \ref{subsec.phase} but is time independent).  The temperature is adjusted to keep the fraction liquid and fraction solid more or less constant.


\begin{table}[ht]
\begin{tabular}{|c||c|ccc|}
\hline
$x_O$&$\bar{\Gamma}$&$x_O^s$&$x_O^l$&$x_O^i$\\
\hline
0.75 &204.5(8)&0.806(1)&0.699(3)&0.740(7)\\
0.50 &221.7(9)&0.552(1)&0.454(3)&0.494(8)\\
0.25 &213.2(7)&0.250(2)&0.249(2)&0.252(6)\\
\hline
\end{tabular}
\caption{Equilibrium compositions of our 55296 ion runs. 
Oxygen number fraction of the whole system is $x_O$, Coulomb parameter averaged over the last third of the run $\bar{\Gamma}$ (determined from the final temperature and $x_O$). 
The composition of the solid is $x^s_O$, the liquid is $x^l_O$, and the interface regions is $x^i_O$.
Statistical errors in the last digit are quoted in parentheses. 
}
\label{tableone}
\end{table}

\begin{figure}[ht]
\begin{center}
\includegraphics[width=3.5in,angle=0,clip=true] {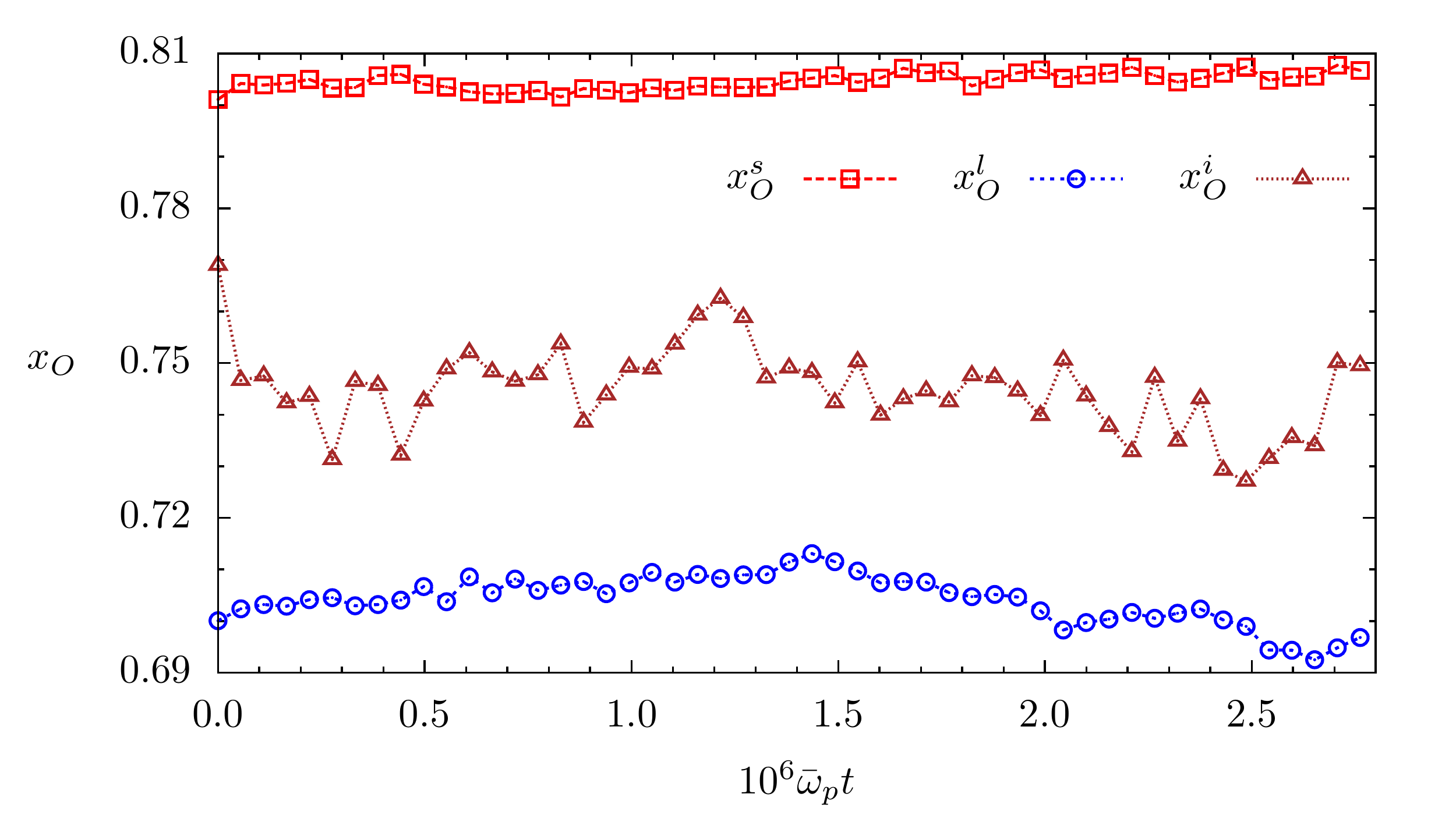}
\caption{(Color on line) Number fraction of oxygen vs simulation time for liquid $x_O^l$ (blue circles), solid $x_O^s$ (red squares) and interface $x_O^i$ (brown triangles) regions.  This is for a 55296 ion simulation that is overall 75\% oxygen and 25\% carbon.}
\label{Fig4}
\end{center}
\end{figure}

In Fig. \ref{Fig4} we show the composition (oxygen number fraction $x_O$) of the solid $x_O^s$, liquid $x_O^l$ and interface $x_O^i$ regions as a function of time. The interface composition is seen to remain near the average value $x_O^i\approx 0.75$ while the solid becomes enriched in oxygen so that $x_O^s\approx 0.8$ and the liquid is depleted in oxygen $x_O^l\approx 0.7$.  The compositions, averaged over the final third of the run, are collected in Table \ref{tableone}.

\begin{figure}[ht]
\begin{center}
\includegraphics[width=3.5in,angle=0,clip=true] {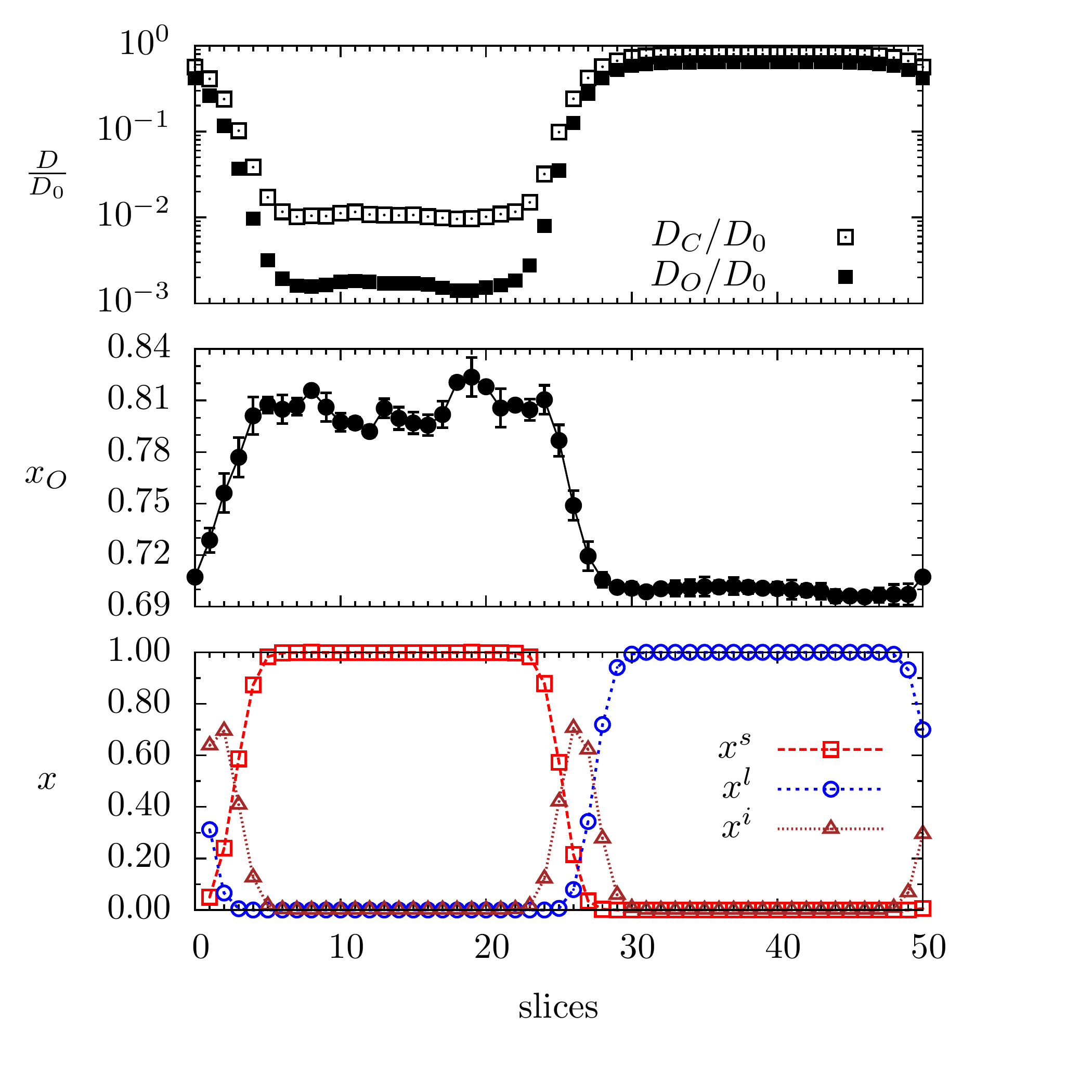}
\caption{(Color on line) Number fraction of oxygen $x_O$ for different slices (regions, see text ) of the simulation volume (middle panel).  Also show in the top panel are diffusion constants $D$ for carbon ions (open squares) and for oxygen ions (filled squares).  Finally the lower panel shows the fraction of ions that are solid (squares), liquid (circles), or interface (triangles).  This is averaged over the last third of the run of the 55296 ion simulation that is overall 75\% oxygen and 25\% carbon.}
\label{Fig5}
\end{center}
\end{figure}

In Fig. \ref{Fig5} we show the oxygen composition $x_O$ vs position in the simulation volume at the end of the run.  We divide up the simulation volume into fifty slices according to the $z$ coordinate, with slice 1 being at the bottom of Fig. \ref{Fig1} and slice 50 being at the top.  For reference we show in Fig. \ref{Fig5} the fraction of ions in a given slice that are in the solid, liquid, and interface regions.  For example, we see that slices 6 to 21 are all solid.  We note that there are some fluctuations in the composition of the solid region, as a function of position, and there are gradients in composition across the interface regions.

We calculated diffusion constants using the methods of ref. \cite{soliddiffusion} in order to check on equilibration.  Fig. \ref{Fig5} shows diffusion constants compared to a reference value $D_0$,
\begin{equation}
D_0=\frac{3\bar\omega_p a^2}{\Gamma^{4/3}},
\label{D_0}
\end{equation}
for both carbon and oxygen ions at various positions in the simulation volume, see also Table \ref{tabletwo}.    Diffusion is seen to be relatively fast in the liquid region with $D_C^l$ for carbon being only slightly larger than $D_O^l$ for oxygen ions.  This is consistent with the results of ref. \cite{liquidnediffusion}.  In contrast $D^s_C$ in the solid is almost 100 times smaller than in the liquid.  This is similar to the one component solid diffusion results of ref. \cite{soliddiffusion}.  Furthermore in the solid, $D^s_O$ is much smaller than $D^s_C$.  The composition of the solid can equilibrate by carbon ions diffusing in to reduce $x_O^s$ or diffusing out to increase $x_O^s$.  One may not need to wait for the oxygen ions to diffuse through out the solid.  Therefore, we expect the equilibration time of our system to be determined by $D^s_C$ in the solid instead of the smaller $D^s_O$.  We find that diffusion is isotropic in the interior of the liquid and solid regions and somewhat non-isotropic near the interfaces.  


\begin{table}[ht]
\begin{tabular}{|c||cc|cc|}
\hline
$x_O$&$D_C^l$&$D_O^l$&$D_C^s$   &$D_O^s$\\
\hline
0.75 &0.80(1)&0.64(1)&0.011(3)  &0.0017(1)\\
0.50 &0.67(1)&0.55(1)&0.0070(4) &0.0012(1)\\ 
0.25 &0.64(1)&0.52(1)&0.0031(2) &0.0007(1)\\
\hline
\end{tabular}
\caption{Diffusion coefficients of liquid and solid phases averaged over the last third of the runs.
Results are expressed as $D_X^p$ in units of $D_0$. 
The letter $p$ denotes the phase, $s$ for solid and $l$ for liquid, while $X$ stands for ion species, $C$ for carbon and $O$ for oxygen. 
Statistical errors in the last digit are quoted in parentheses. 
}
\label{tabletwo}
\end{table}

\subsection{Run with 50\% oxygen}
\label{subsec.O50}

\begin{figure}[ht]
\begin{center}
\includegraphics[width=3.5in,angle=0,clip=true] {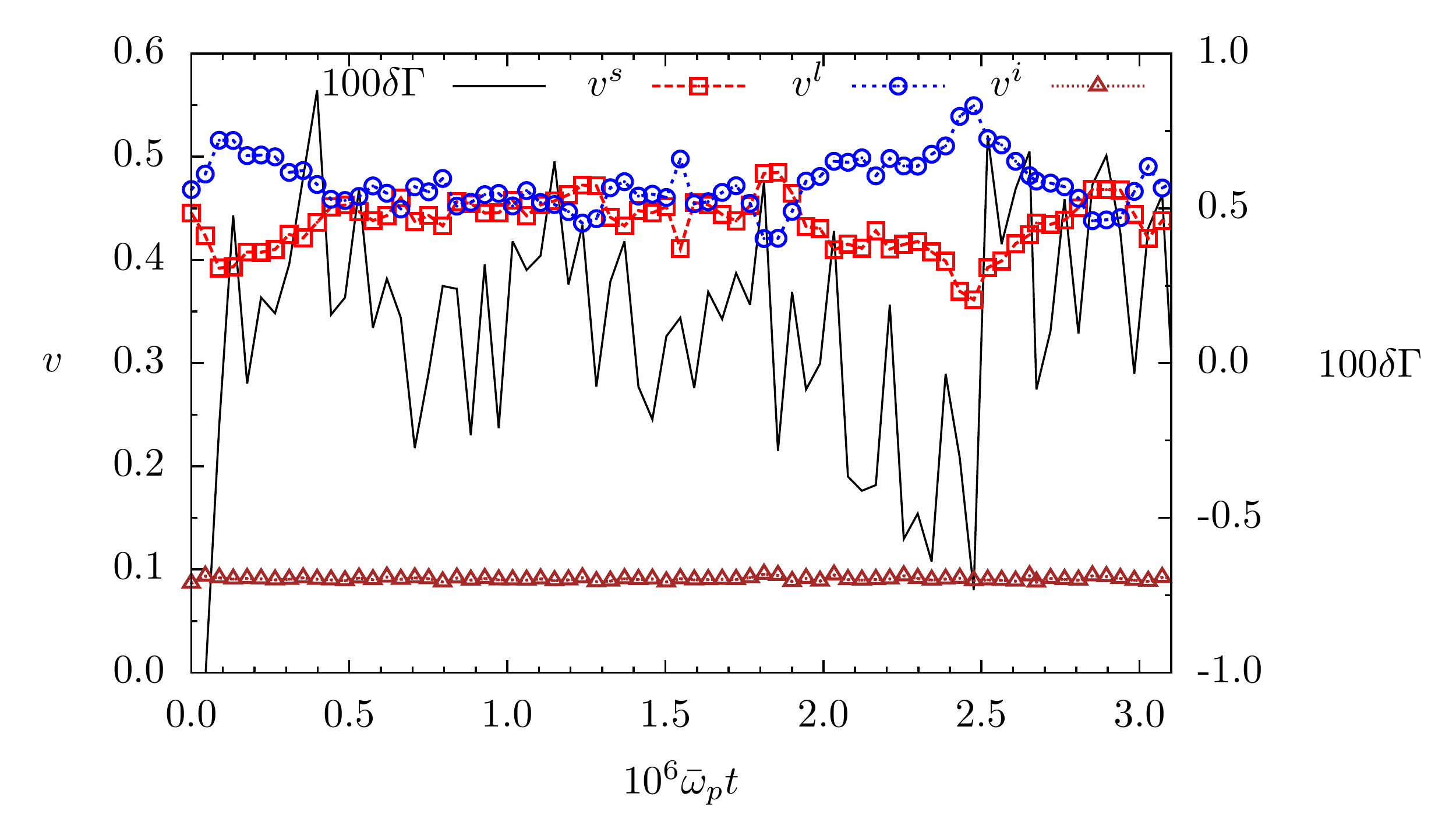}
\caption{(Color on line) Fractional fluctuations in Coulomb parameter $\delta\Gamma$ (solid black line and right hand scale), see Eq. \ref{deltaGamma} vs simulation time $t$ for a 55296 ion system with 50\% oxygen and 50\% carbon.  Also shown are the fraction of the system that is solid  (red squares), liquid (blue circles) or interface (brown triangles).}
\label{Fig8}
\end{center}
\end{figure}

\begin{figure}[ht]
\begin{center}
\includegraphics[width=3.5in,angle=0,clip=true] {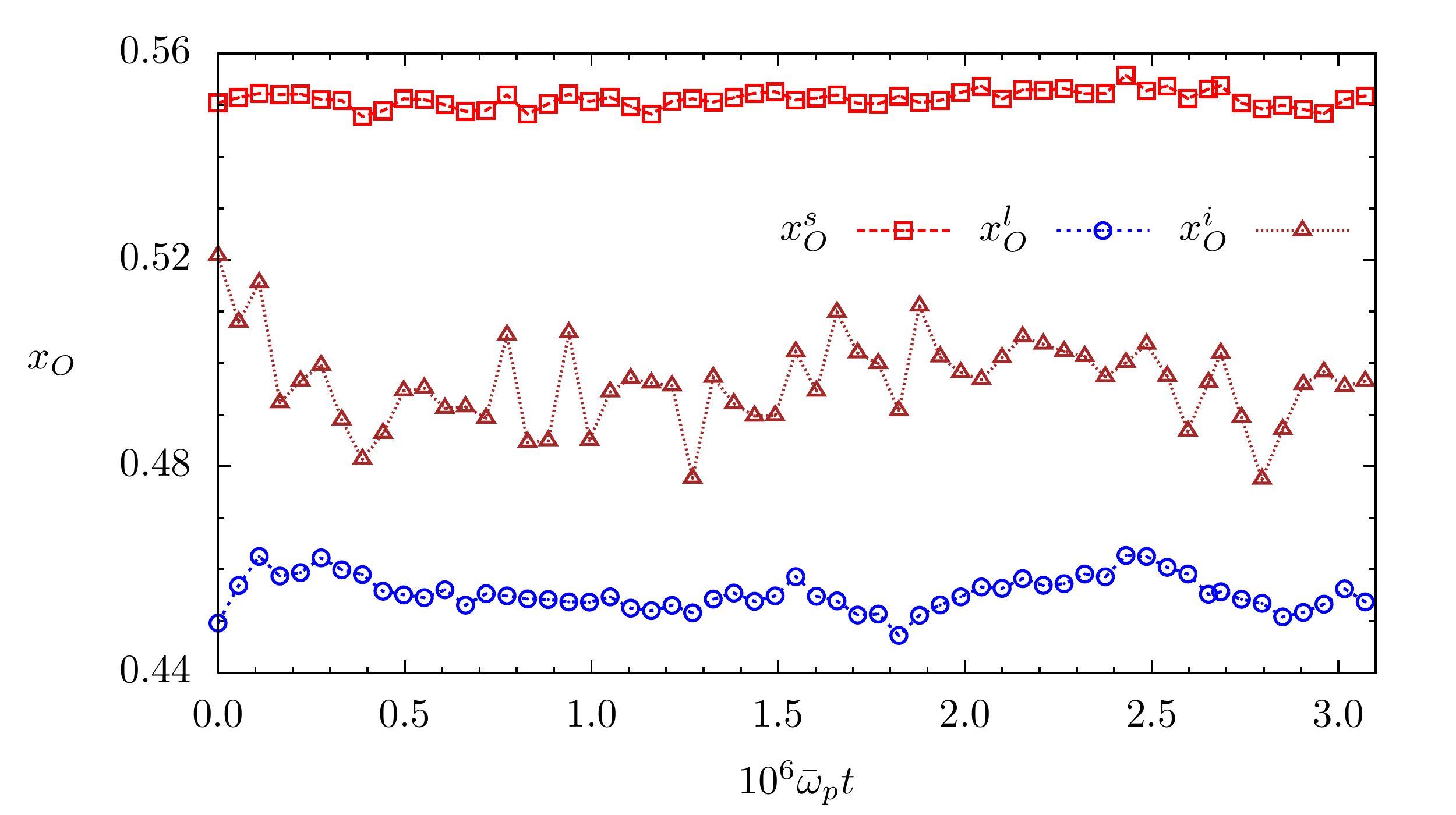}
\caption{(Color on line) Number fraction of oxygen vs simulation time for liquid $x_O^l$ (blue circles), solid $x_O^s$ (red squares) and interface $x_O^i$ (brown triangles) regions.  This is for a 55296 ion simulation that is overall 50\% oxygen and 50\% carbon.}
\label{Fig9}
\end{center}
\end{figure}

Our second 55296 ion run has an overall composition of 50\% oxygen and 50\% carbon.  We started it in a very similar manner as for the 75\% oxygen run except that the initial composition of the solid was assumed to be 55\% oxygen, while the liquid was assumed to be 45\% oxygen.  This is based on our earlier results \cite{WDPRL}.  Figure \ref{Fig8} shows the fluctuations in $\Gamma$, and the fractions of the system that are liquid, solid, and interface as a function of time.  The compositions of these regions vs time are shown in Fig. \ref{Fig9} and the compositions averaged over the final third of the run are collected in Table \ref{tableone}.   We note that there is very little time dependence to these compositions.

\begin{figure}[ht]
\begin{center}
\includegraphics[width=3.5in,angle=0,clip=true] {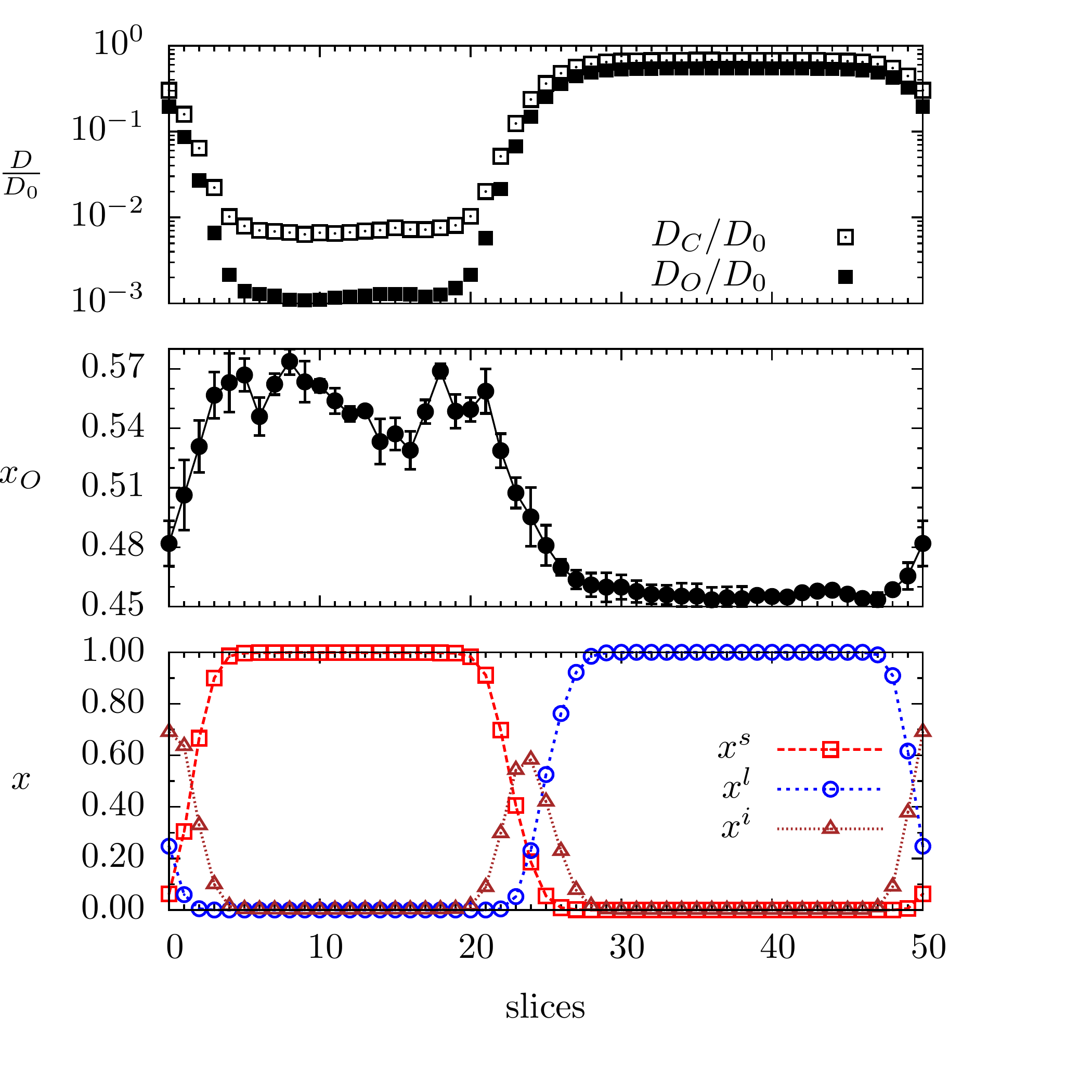}
\caption{Color on line) Number fraction of oxygen $x_O$ for different slices (regions, see text ) of the simulation volume (middle panel).  Also show in the top panel are diffusion constants $D$ for carbon ions (open squares) and for oxygen ions (filled squares).  Finally the lower panel shows the fraction of ions that are solid (squares), liquid (circles), or interface (triangles).  This is averaged over the last third of the run of the 55296 ion simulation that is overall 50\% oxygen and 50\% carbon.}
\label{Fig10}
\end{center}
\end{figure}

In Fig. \ref{Fig10} we show composition vs position.  Again there are some fluctuations in the composition of the solid.  Diffusion constants $D$ are collected in Table \ref{tabletwo}.  In general $D$ as a function of position is very similar to our results from the 75\% oxygen simulation.  However we note that $D$ in the solid is somewhat smaller than for the 75\% oxygen run.

\subsection{Run with 25\% oxygen}
\label{subsec.O25}

\begin{figure}[ht]
\begin{center}
\includegraphics[width=3.5in,angle=0,clip=true] {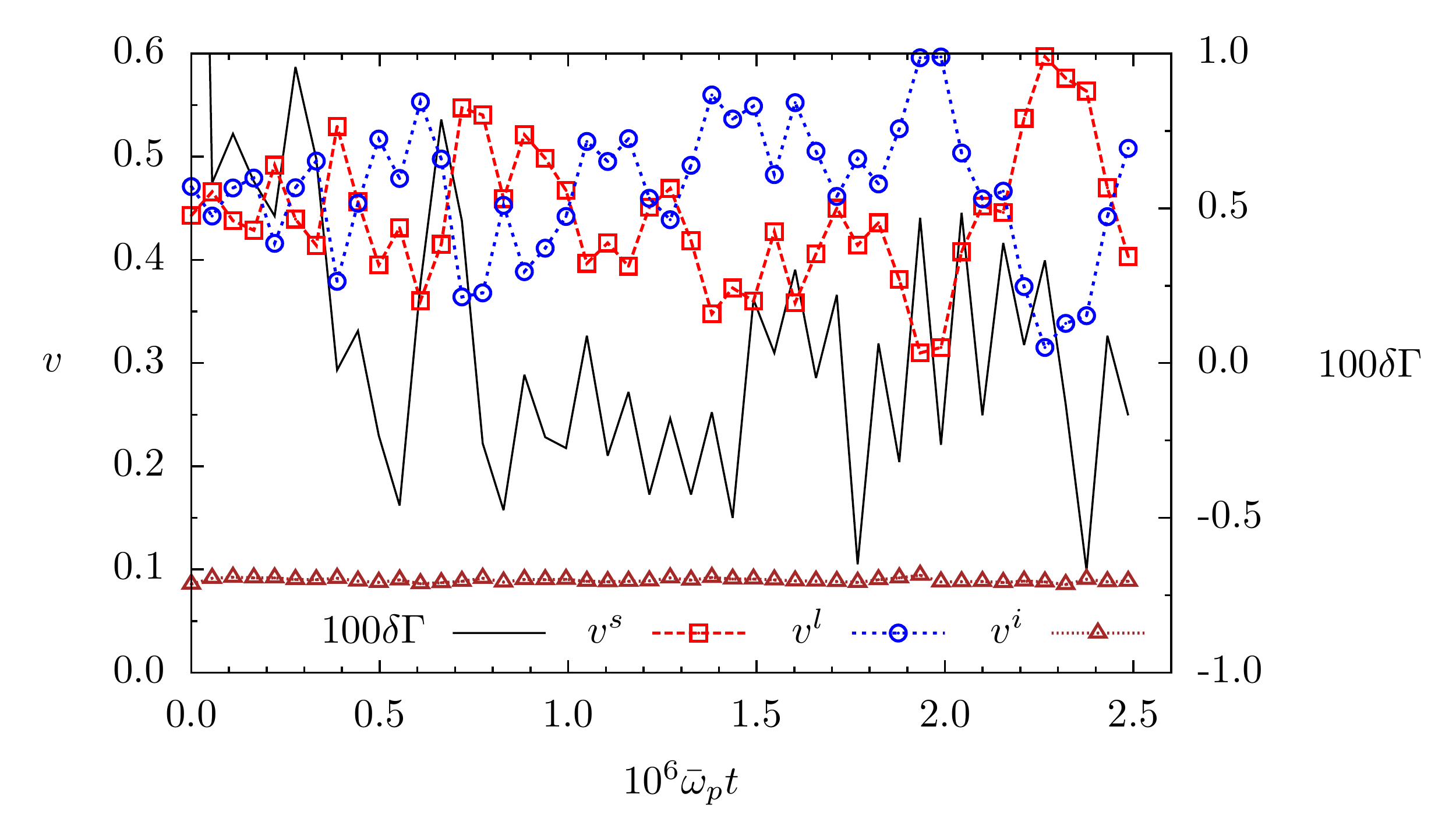}
\caption{(Color on line) Fractional fluctuations in Coulomb parameter $\delta\Gamma$ (solid black line and right hand scale), see Eq. \ref{deltaGamma} vs simulation time $t$ for a 55296 ion system with 25\% oxygen and 75\% carbon.  Also shown are the fraction of the system that is solid  (red squares), liquid (blue circles) or interface (brown triangles).}
\label{Fig12}
\end{center}
\end{figure}

\begin{figure}[ht]
\begin{center}
\includegraphics[width=3.5in,angle=0,clip=true] {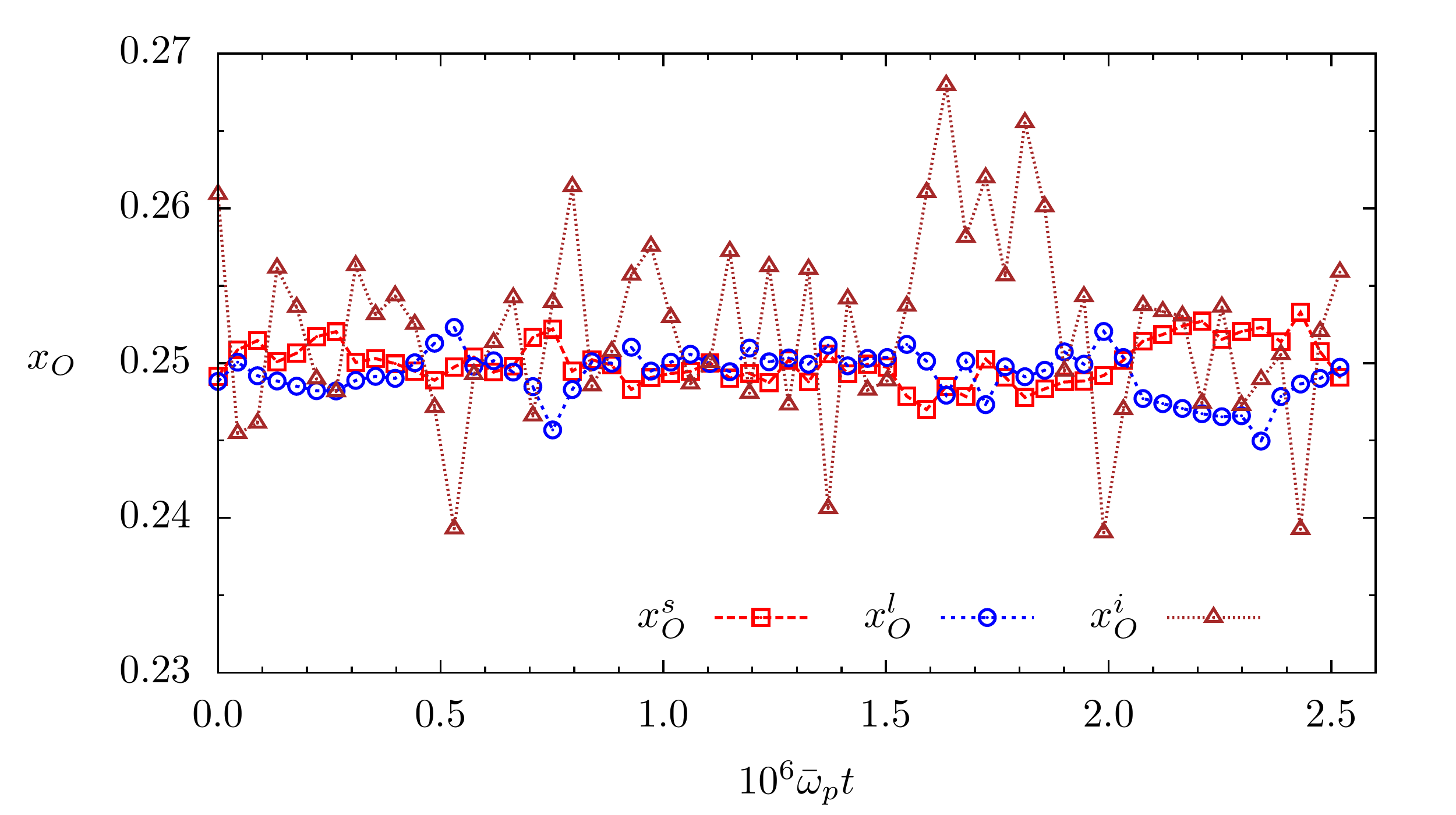}
\caption{(Color on line) Number fraction of oxygen vs simulation time for liquid $x_O^l$ (blue circles), solid $x_O^s$ (red squares) and interface $x_O^i$ (brown triangles) regions.  This is for a 55296 ion simulation that is overall 25\% oxygen and 75\% carbon.}
\label{Fig13}
\end{center}
\end{figure}

Our final 55296 ion run has an overall composition of 25\% oxygen and 75\% carbon.  We started it in a very similar manner as for the 75\% oxygen run except that the initial composition of the solid was assumed to be equal to that of the liquid, with both at 25\% oxygen.  This is because our earlier simulations found only small chemical separation \cite{WDPRL}.  Figure \ref{Fig12} shows the fluctuations in $\Gamma$, and the fractions of the system that are liquid, solid, and interface as a function of time.  The compositions of these regions vs time are shown in Fig. \ref{Fig13} and the compositions averaged over the final third of the run are collected in Table \ref{tableone}.   We note that all compositions are near 25\% oxygen.  However, there is a small tendency for the interface regions to be slightly enriched in oxygen compared to both the liquid and solid regions.  Note that the fluctuations in the interface composition are larger than the fluctuations in the compositions of the liquid and solid regions because the interface contains fewer ions.  The composition of the liquid is very close to that of the solid although, at some times there are small fluctuations in these compositions.  Of course in the thermodynamic limit the composition of the interface is not relevant.  Nevertheless, our MD simulation could be showing a real effect where the interface is enriched in oxygen even if the liquid and solid have nearly equal compositions.

\begin{figure}[ht]
\begin{center}
\includegraphics[width=3.5in,angle=0,clip=true] {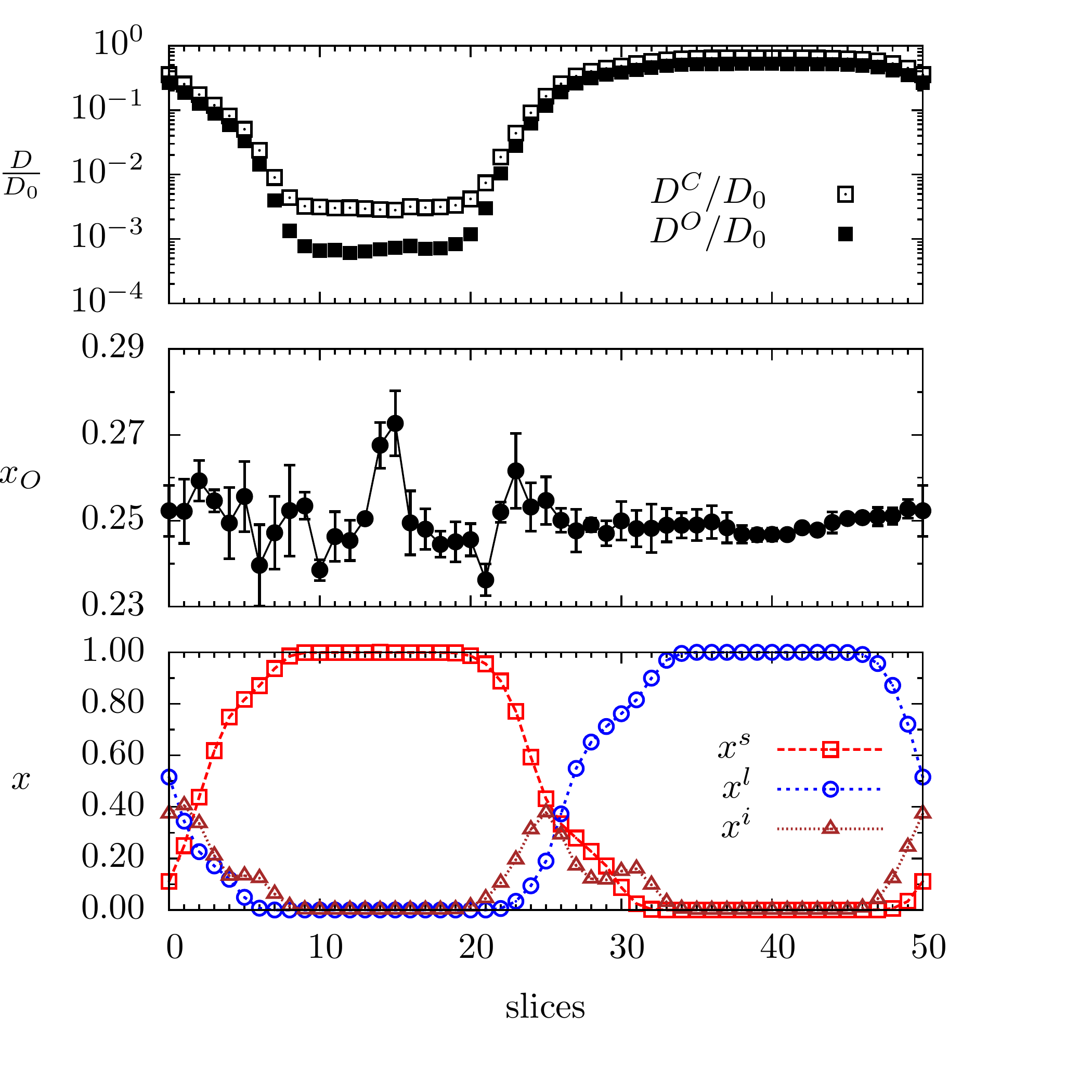}
\caption{Color on line) Number fraction of oxygen $x_O$ for different slices (regions, see text ) of the simulation volume (middle panel).  Also show in the top panel are diffusion constants $D$ for carbon ions (open squares) and for oxygen ions (filled squares).  Finally the lower panel shows the fraction of ions that are solid (squares), liquid (circles), or interface (triangles).  This is averaged over the last third of the run of the 55296 ion simulation that is overall 25\% oxygen and 75\% carbon.}
\label{Fig14}
\end{center}
\end{figure}

In Fig. \ref{Fig14} we show composition vs position.  The composition is nearly constant independent of position.  However, again there are some fluctuations in the composition of the solid.  Diffusion constants $D$ are collected in Table \ref{tabletwo}.  Now $D$ in the solid, for both carbon and oxygen, are smaller than in the 75\% or 50\% oxygen runs.

\subsection{Carbon - Oxygen Phase diagram}
\begin{figure}[ht]
\begin{center}
\includegraphics[width=3.5in,angle=0,clip=true] {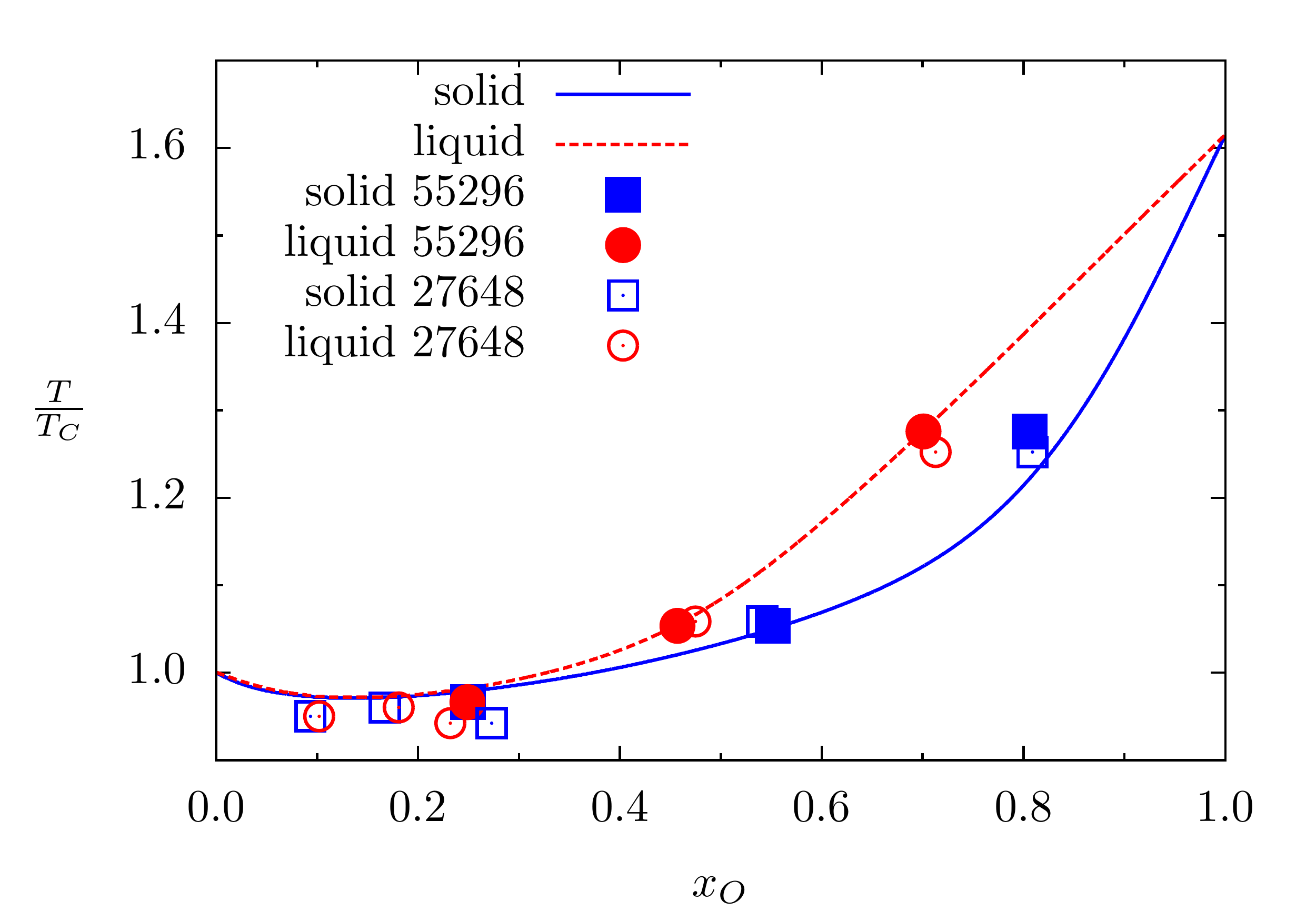}
\caption{Carbon-oxygen phase diagram plotting the composition of the liquid phase (upper red curve or circles) that is in equilibrium with the solid phase (lower blue curve or squares).  Our present results from 55296 ion simulations are filled symbols while the open symbols are our previous results with 27648 ions from ref. \cite{WDPRL}.  The curves are the results of Medin and Cumming \cite{medin}. }
\label{Fig15}
\end{center}
\end{figure}

We now present the liquid-solid phase diagram implied by the simulations in Sections \ref{subsec.O75}, \ref{subsec.O50}, and \ref{subsec.O25}.  Figure \ref{Fig15} shows the phase diagram as a function of $x_O$.  The $y$ axis is the melting temperature $T$ divided by the melting temperature $T_C$ for pure carbon.  We assume the pure carbon system melts at $\Gamma_m=178.4$ \cite{WDPRL}.  This differs slightly from the one component plasma result because we include the effects of electron screening.

The points plotted in Fig. \ref{Fig15} are listed in Table \ref{tableone}.  Overall our 55296 ion results are close to our previous results that used 27648 ion simulations \cite{WDPRL}.  However, there are some noticeable differences.  The 25\% oxygen simulation with 55296 ions has nearly equal liquid and solid compositions, while the previous 27648 ion simulation found the solid slightly enriched in oxygen and found a slightly lower melting temperature.  The 50\% oxygen simulation with 55296 ions has a slightly larger difference in composition between the liquid and solid than previous 27648 ion simulations.  Finally the 75\% oxygen simulation with 55296 ions has a slightly higher melting temperature than the previous 27648 ion result.   This suggests that finite size effects, while not zero for the 27648 ion simulations, are relatively small.

The agreement between our 55296 ion results and the model of Medin and Cumming is excellent.  All three liquid compositions, and two of the three solid compositions, from Table \ref{tableone} are very close to Medin and Cumming's results.  The only small difference is that our solid composition from the 75\% oxygen simulation is not quite as oxygen rich as Medin and Cumming.  This one difference could be due to a small remaining systematic error in our simulation, or to a small error in the model liquid and solid free energies used by Medin and Cumming.  In any case, the small finite size corrections in going from 27648 ion to 55296 ion simulations improve the agreement between our MD simulations and Medin and Cumming.  

What are the nature of the errors from finite size and or non-equilibrium effects?  Although small, the errors appear to go in different directions for our three simulations.  For the 25\% oxygen run the 55296 ion simulation has a higher melting temperature and much smaller difference between the composition of the liquid and solid compared to 27648 ion results.  This is in a region of the phase diagram where the melting temperature is almost independent of composition.  Therefore the equilibrium compositions may be very sensitive to any small errors.

For 50\% oxygen, the 55296 ion simulation has a larger difference between the compositions of the liquid and solid compared to 27648 ion results.  Perhaps the simplest finite size effect would arise if the composition gradient across an interface extended over a distance comparable to the box size.  In this case simulations with small boxes might have more nearly equal liquid and solid compositions than simulations with larger boxes.   For example in Fig. \ref{Fig12} the gradient in composition extends over a distance up to perhaps as many as 10 slices and the distance between the two liquid-solid interfaces is 25 slices.  In contrast, the 27648 ion runs from ref. \cite{WDPRL} are in a cubical box where the distance between the two liquid-solid interfaces is a factor of two smaller (equivalent to of order 12.5 slices).  Thus, there could be some small finite size effects for 27648 ion runs.   

Finally, there could be a statistical component to the errors coming from a variety of non-equilibrium effects.  For example fluctuations in the location of an interface could create a new solid region and the composition of this region might not have time to equilibrate.   Alternatively there could be composition changes from large fluctuations.  One could test for a variety of errors of this type by simply repeating these simulations a number of times with different initial conditions.  Unfortunately, we have not had time to do this for the present paper.  However, we plan to do this in the future.

The good agreement between our phase diagram and that of Medin and Cumming strongly suggest that the remaining errors in our direct MD simulation approach are small.  And in addition, it suggests that the model free energies employed by Medin and Cumming are good, at least for the carbon-oxygen system.   Furthermore, {\it we conclude that the phase diagram for the carbon-oxygen system is accurately known.} We emphasize that our direct MD simulations only work at all because diffusion in the solid phase is relatively fast \cite{soliddiffusion}.  Had diffusion been slow then it would be very difficult to equilibrate the solid phases.  

We find direct two-phase MD simulations can accurately determine liquid-solid phase equilibria.  This result is very useful because direct MD simulation can be applied to many other systems, including very complex ones.  Furthermore direct MD simulations for a few compositions may provide very helpful benchmarks for simpler models.  Note that simulations with a somewhat large number of particles may be necessary and these may have to be run for extended simulation times.  However, rapid advances in computer power should make such simulations even easier in the future.

\section{Results for Oxygen and Selenium Systems}
\label{results_oxygen_selenium}
In this section we present results for the phase diagram and diffusion constants for oxygen and selenium systems.  This two-component system has a much larger ratio of charges, and this leads to a very different phase diagram, than for the carbon and oxygen system of Sec. \ref{CO}.  We perform MD simulations with both 27648 and 55296 ions in order to monitor finite size effects.  In general we present figures for the larger 55296 ion simulations and then include both 27648 and 55296 ion results in tables.  These simulations follow closely the formalism of Sec. \ref{Formalism} and the procedures of Sec. \ref{CO}.  

However the interface finding algorithm in Sec. \ref{subsec.phase} is slightly modified.  The bond-angles for oxygen ions can be liquid-like (more random) even in the solid phase.  Therefore, we only take into account selenium ions in the bond angle algorithm.  By ignoring all oxygen ions in the system we determine whether a selenium ion is solid-like or liquid-like by analyzing how it bonds to its neighboring selenium ions.  As in Sec. \ref{subsec.phase}, ions are considered neighbors if they are within a distance $r_{ij}<4a$ of each other. Once all selenium ions have been identified as solid-like or liquid-like we tag an ion (oxygen or selenium) as being in the bulk of the solid or liquid if a large majority ($>80\%$) of the selenium ions within $4a$ of it are the same phase. If this criterion is not met, then the ion is determined to be in the interface.

We now discuss runs with 98\%, 90\%, 80\%, 70\%, 60\%, and 50\% selenium and then we will collect results for the oxygen-selenium phase diagram.  Although most runs appear to be equilibrated by the end of the simulations, we note that the 60\% and 50\% selenium runs are not equilibrated after using a reasonable amount of computer time.  We discuss this below. 

\begin{figure}[ht]
\begin{center}
\includegraphics[width=3.5in,angle=0,clip=true] {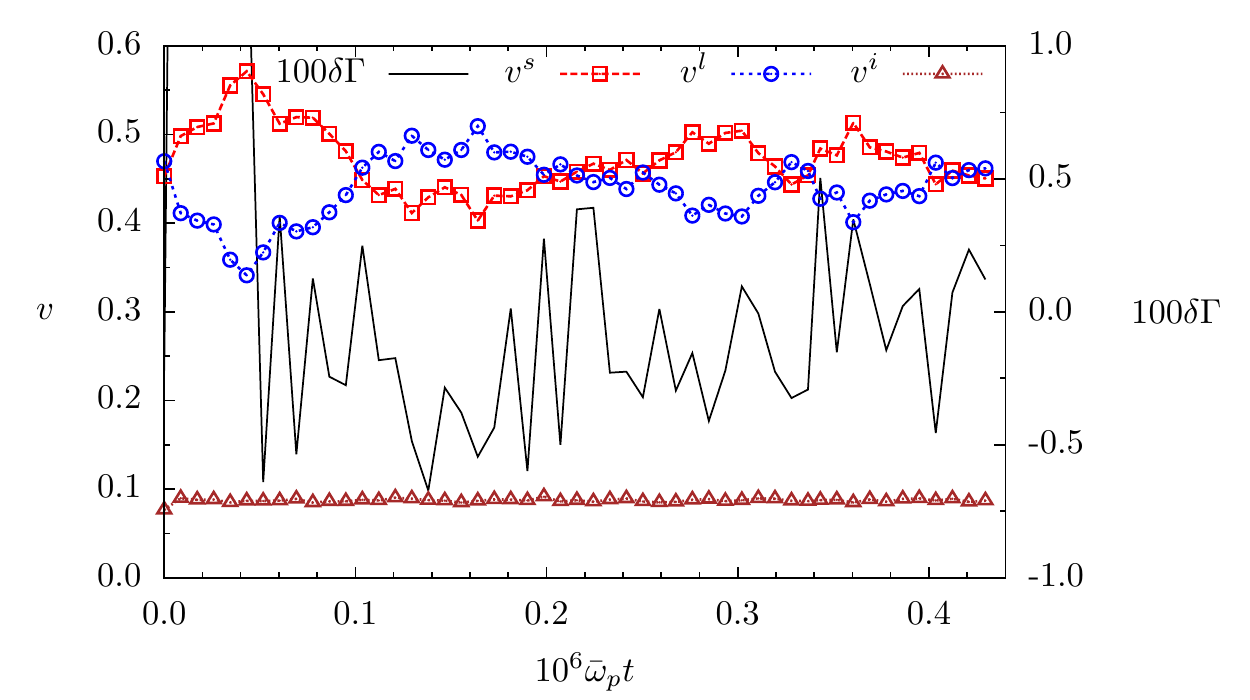}
\caption{(Color on line) Volume fractions of solid (squares), liquid (circles), and interface (triangles) versus time for a 55296 ion simulation that is overall 2\% oxygen and 98\% selenium.  Also shown are fluctuations $\delta\Gamma$ in the Coulomb parameter (solid line), see Eq. \ref{deltaGamma}.}
\label{Fig15a}
\end{center}
\end{figure}

\begin{figure}[ht]
\begin{center}
\includegraphics[width=3.5in,angle=0,clip=true] {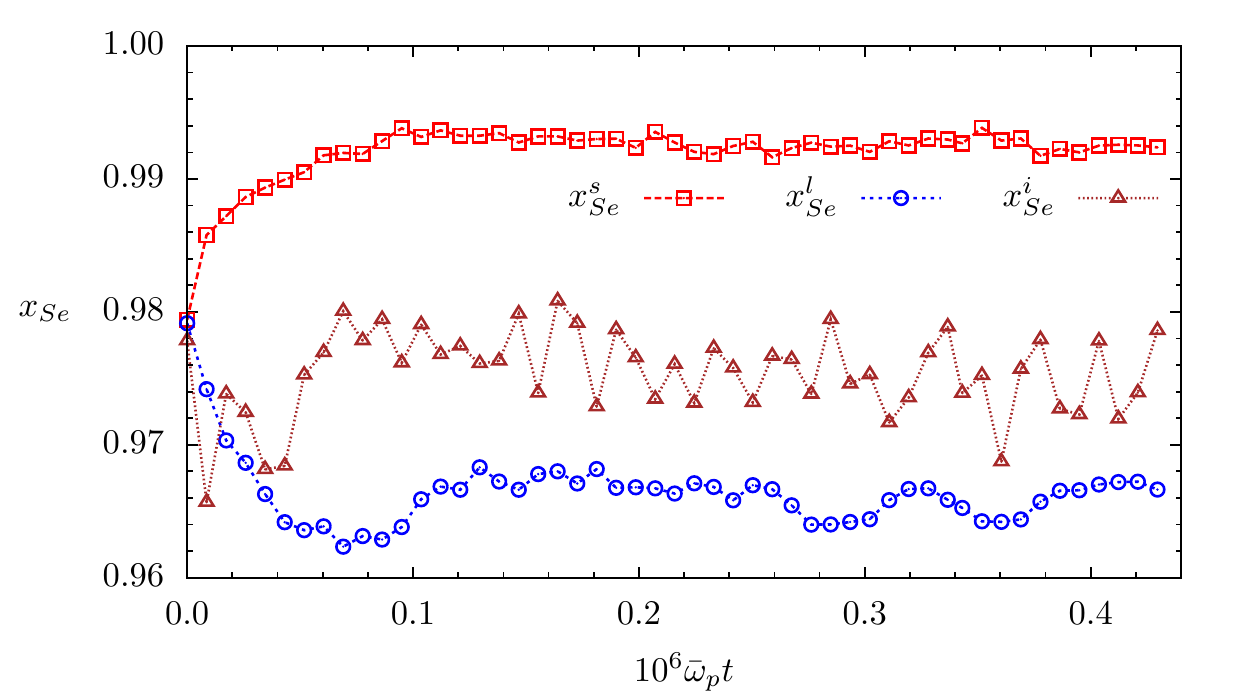}
\caption{(Color on line) Number fraction of selenium $x_{Se}$ in the solid (squares), liquid (circles), and interface (triangles) versus time for a 55296 ion simulation that is overall 2\% oxygen and 98\% selenium.}
\label{Fig16}
\end{center}
\end{figure}

\subsection{Run with 98\% Selenium}
\label{subsec.Se98}

As for the carbon-oxygen systems we prepare solid and liquid initial conditions separately and then combine them to obtain the full 55296 ion initial conditions.  

To prepare the solid we start with a 432 ion system that is composed of 98\% selenium ions with random positions and velocities.  The system is evolved at a temperature close to its expected melting temperature for $t\bar\omega_p=4500$ using a time step of $0.11\bar\omega_p$. Due to finite size effects it solidifies.  We then make 8 copies of this 432 ion system to obtain a 3456 ion solid and evolved it for $t\bar\omega_p\simeq20000$ keeping a time step of $0.11\bar\omega_p$.  Eight copies of this system are made to obtain a 27648 ion solid. This solid is evolved for $t\bar\omega_p\simeq20000$ with a time step of $0.22\bar\omega_p$. 

The liquid is prepared in a similar fashion. We start with a 3456 ion system that is 98\% selenium ions with random positions and velocities.  This system is evolved for $t\bar\omega_p\simeq20000$ with a time step of $0.11\bar\omega_p$.  Eight copies of this system are made to obtain a 27648 ion liquid which is evolved for $t\bar\omega_p\simeq20000$ using a time step of $0.22\bar\omega_p$.  Finally, we place the 27648 ion liquid configuration on top of the 27648 ion solid configuration to form the full 55296 ion initial conditions.

The temperature was adjusted during the run to keep approximately equal volume fractions of solid and liquid as shown in Fig. \ref{Fig15a}.   The number fraction of selenium $x_{Se}$ in the solid, liquid and interface are shown in Fig. \ref{Fig16} versus simulation time, see also Table \ref{tablethree}.  These fractions do not change much in the second half of the run suggesting that the system is (at least approximately) equilibrated.

\begin{table}[ht]
\begin{tabular}{|c||c|ccc|}
\hline
$x_{Se}$&$\bar{\Gamma}$&$x_{Se}^s$&$x_{Se}^l$&$x_{Se}^i$\\
\hline
0.98 &201.1(6)&0.9926(5)&0.966(1)&0.975(2)\\
0.90 &213.4(8)&0.9708(7)&0.843(1)&0.884(4)\\
0.80 &252.5(6)&0.968(1) &0.681(1)&0.762(7)\\
0.70 &289(1)  &0.970(1) &0.547(1)&0.655(4)\\
0.60 &391(1)  &0.935(2) &0.408(1)&0.513(8)\\
0.50 &459(2)  &0.902(2) &0.308(1)&0.436(9)\\
\hline
\end{tabular}
\caption{Equilibrium compositions of our 55296 ion runs. 
Selenium number fraction of the whole system is $x_{Se}$, Coulomb parameter averaged over the last third of the run $\bar{\Gamma}$ (determined from the final temperature and $x_O$). 
The composition of the solid is $x_s$, the liquid is $x_l$, and the interface regions is $x_i$.
Statistical errors are quoted in the last digit in parentheses.}
\label{tablethree}
\end{table}

\begin{figure}[ht]
\begin{center}
\includegraphics[width=3.5in,angle=0,clip=true] {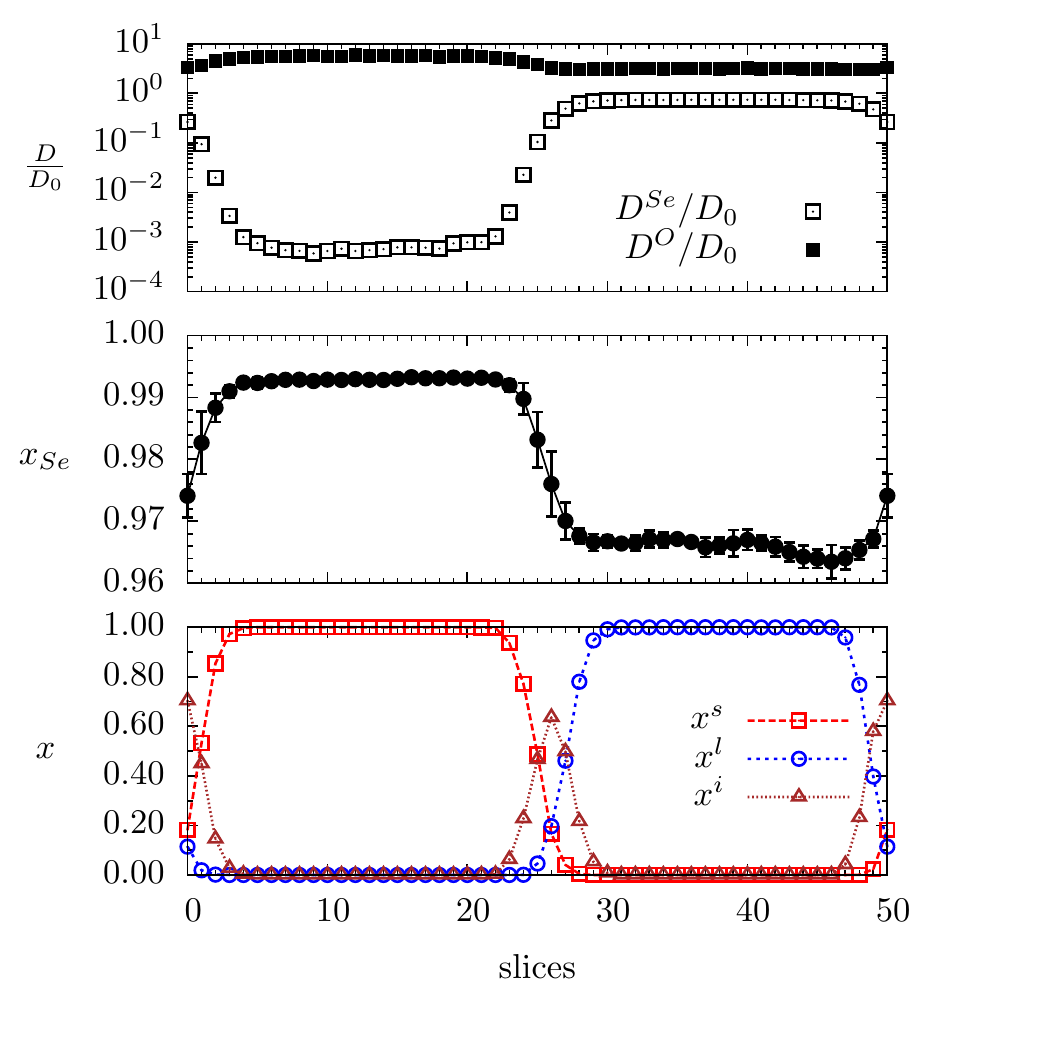}
\caption{(Color on line) Diffusion constant $D$ (top panel)) for selenium (open squares) and oxygen ions (filled squares) for different slices (regions) of the simulation volume. Number fraction of selenium $x_{Se}$ (middle panel) for different slices.  Fraction of ions $x$ (lower panel) that are in the solid (squares), liquid (circles) and interface (triangles)for different slices.  This is an average over the last $1.4\times 10^5\bar\omega_pt$ of a 55296 ion simulation that is overall 2\% oxygen and 98\% selenium.}
\label{Fig17}
\end{center}
\end{figure}

In Fig. \ref{Fig17}, in the middle panel, we show the number fraction of selenium $x_{Se}$ for different slices (regions) of the simulation volume.  This is at the end of the simulation.  In the solid, $x_{Se}$ is seen to be nearly constant and independent of position.  This is consistent with the system having reached equilibration.   Diffusion constants $D$ are also shown in Fig. \ref{Fig17} in the top panel and listed in Table \ref{tablefour}.  For selenium $D$ in the solid is seen to be three orders of magnitude smaller than $D$ in the liquid.  However the behavior is very different for oxygen.  Diffusion is nearly the same in the liquid and solid regions!  Indeed in the solid, $D$ for oxygen is over three orders of magnitude larger than $D$ for selenium.   Presumably the effective size of oxygen ions in the solid (ion sphere radius) is small enough so that the oxygen can diffuse relatively easily through the larger crystal lattice of selenium ions.  Note that this behavior for oxygen in a selenium crystal is very different from that for carbon in an oxygen crystal as found in Sec. \ref{CO}, see Fig. \ref{Fig5}.  

\begin{table}[ht]
\begin{tabular}{|c||cc|cc|}
\hline
$x_{Se}$&$D_O^l$&$D_{Se}^l$&$D_O^s$   &$D_{Se}^s$\\
\hline
0.98 &3.17(1)&0.736(3)&5.66(1) &0.00080(2)\\
0.90 &3.04(1)&0.729(2)&5.34(3) &0.00064(7)\\
0.80 &2.88(1)&0.721(6)&4.50(9) &0.00044(7)\\
0.70 &2.81(1)&0.771(2)&1.82(1) &0.00014(1)\\
0.60 &2.36(1)&0.617(6)&0.18(2) &0.00002(1)\\
0.50 &2.12(1)&0.547(2)&0.1(1)  &0.00001(1)\\
\hline
\end{tabular}
\caption{Diffusion coefficients of liquid and solid phases averaged over the last third of the runs.
Results are expressed as $D_X^p$ in units of $D_0$. 
The letter $p$ denotes the phase, $s$ for solid and $l$ for liquid, while $X$ stands for ion species, $O$ for oxygen and $Se$ for Selenium. 
Statistical errors are quoted in the last digit in parentheses.}
\label{tablefour}
\end{table}

\subsection{Run with 90\% Selenium}
\label{subsec.Se90}
The inital conditions for the 90\% selenium simulations were prepared similarly to the 98\% ones. The only difference was that the initial compositions of the 432 ion solid and 3456 ion liquid systems were set to 90\% selenium.  We then follow the procedure laid out in Sec. \ref{subsec.Se98} to obtain a 55296 ion configuration.

\begin{figure}[ht]
\begin{center}
\includegraphics[width=3.5in,angle=0,clip=true] {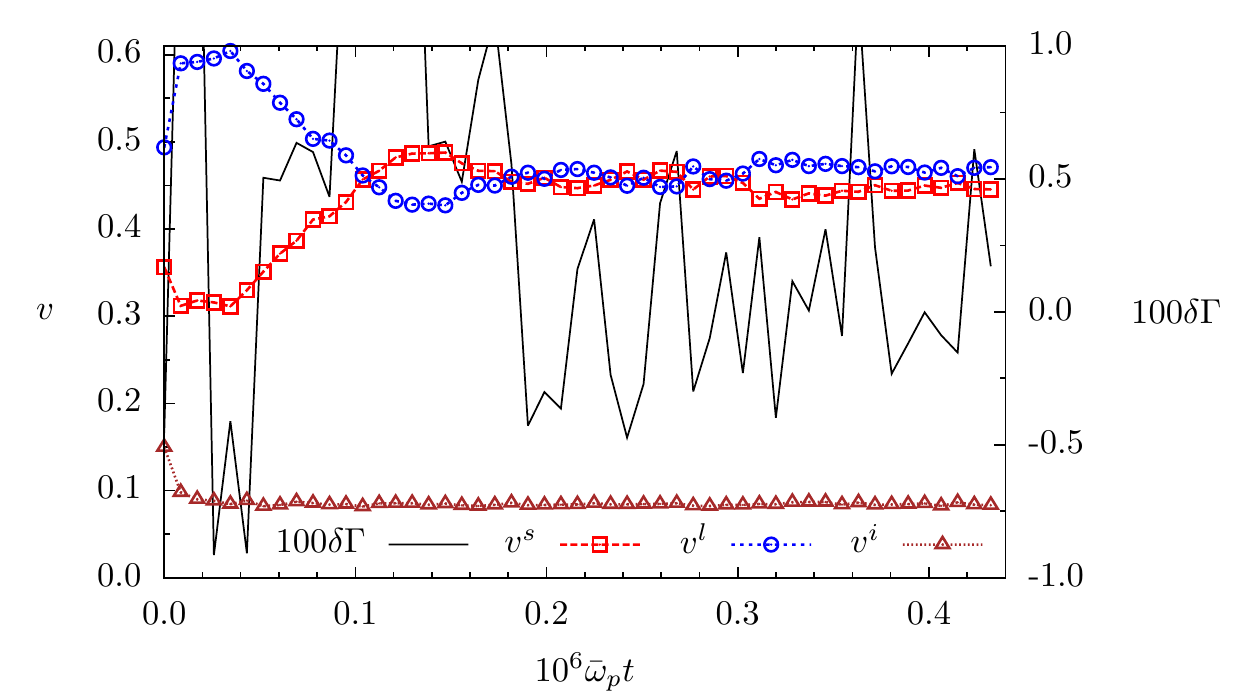}
\caption{(Color on line) Volume fractions of solid (squares), liquid (circles), and interface (triangles) versus time for a 55296 ion simulation that is overall 10\% oxygen and 90\% selenium.  Also shown are fluctuations $\delta\Gamma$ in the Coulomb parameter (solid line), see Eq. \ref{deltaGamma}.}
\label{Fig18}
\end{center}
\end{figure}

\begin{figure}[ht]
\begin{center}
\includegraphics[width=3.5in,angle=0,clip=true] {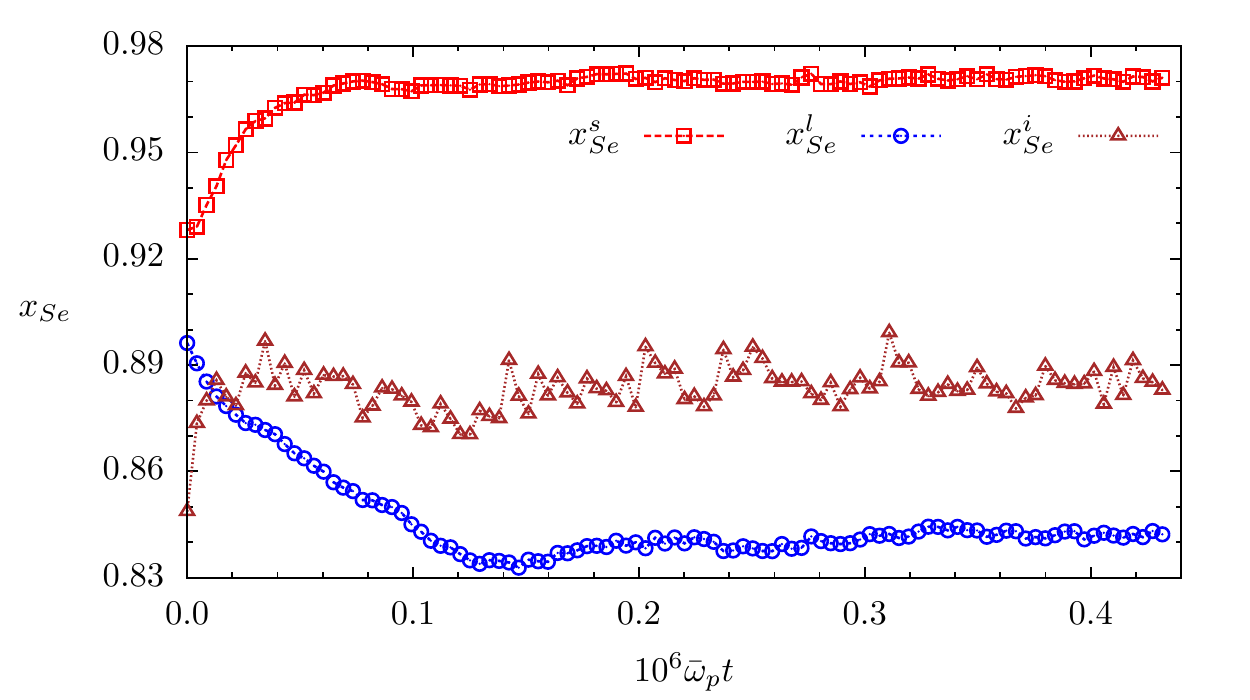}
\caption{(Color on line) Number fraction of selenium $x_{Se}$ in the solid (squares), liquid (circles), and interface (triangles) versus time for a 55296 ion simulation that is overall 10\% oxygen and 90\% selenium.}
\label{Fig19}
\end{center}
\end{figure}

The temperature was adjusted during the run to keep approximately equal volume fractions of solid and liquid as shown in  Fig. \ref{Fig18}.   The number fraction of senium $x_{Se}$ in the solid, liquid and interface are shown in Fig. \ref{Fig19} versus simulation time.  These fractions do not change much in the second half of the run suggesting that the system is (at least approximately) equilibrated, as for the run with 98\% selenium. 

\begin{figure}[ht]
\begin{center}
\includegraphics[width=3.5in,angle=0,clip=true] {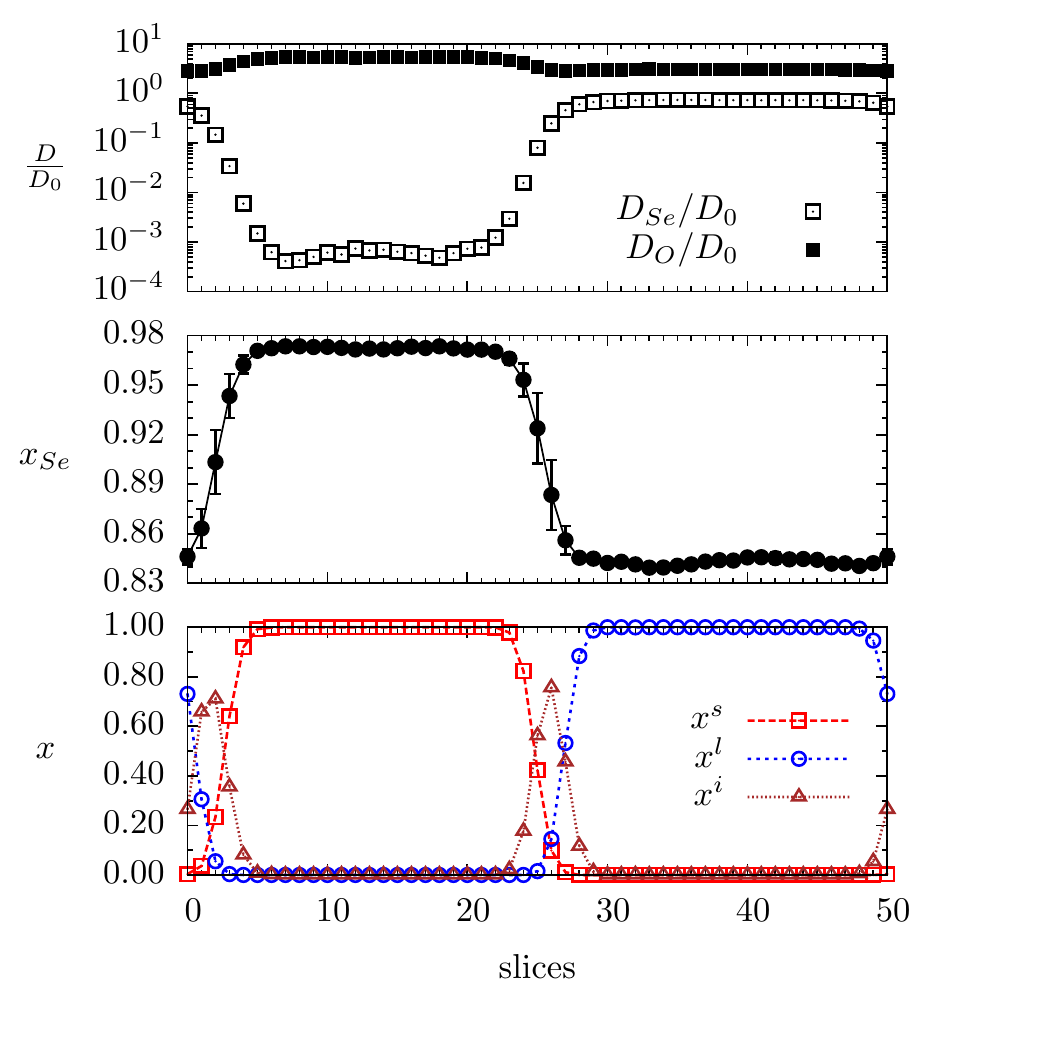}
\caption{(Color on line) Diffusion constant $D$ (top panel)) for selenium (open squares) and oxygen ions (filled squares) for different slices (regions) of the simulation volume. Number fraction of selenium $x_{Se}$ (middle panel) for different slices.  Fraction of ions $x$ (lower panel) that are in the solid (squares), liquid (circles) and interface (triangles)for different slices.  This is an average over the last $1.4\times 10^5\bar\omega_pt$ of a 55296 ion simulation that is overall 10\% oxygen and 90\% selenium.}
\label{Fig20}
\end{center}
\end{figure}

In Fig. \ref{Fig20}, in the middle panel, we show the number fraction of selenium $x_{Se}$ for different slices (regions) of the simulation volume.  This is at the end of the simulation.  In the solid, $x_{Se}$ is seen to be nearly constant and independent of position.  This is consistent with the system having reached equilibration.   Diffusion constants $D$ are also shown in Fig. \ref{Fig20} in the top panel and are seen to behave in a similar way to Fig. \ref{Fig17}. 

\subsection{Run with 80\% Selenium}
\label{subsec.Se80}
The inital conditions for this simulation were prepared similarly to the initial conditions for the 98\% and 90\% selenium systems.  However, expecting the solid to be enriched in selenium and desiring the system to reach equilibrium faster, we start the liquid and the solid subsystems with different initial compositions.  We start with a 432 ion solid configuration that is 90\% selenium and a 3456 liquid configuration that is 70\% selenium.  We then follow the procedure of Sec. \ref{subsec.Se98} to obtain the 55296 80\% selenium system.  We note here that due to their different initial compositions, the solid and the liquid initally have different electron densities.  However the electron densities quickly equilibrate.

\begin{figure}[ht]
\begin{center}
\includegraphics[width=3.5in,angle=0,clip=true] {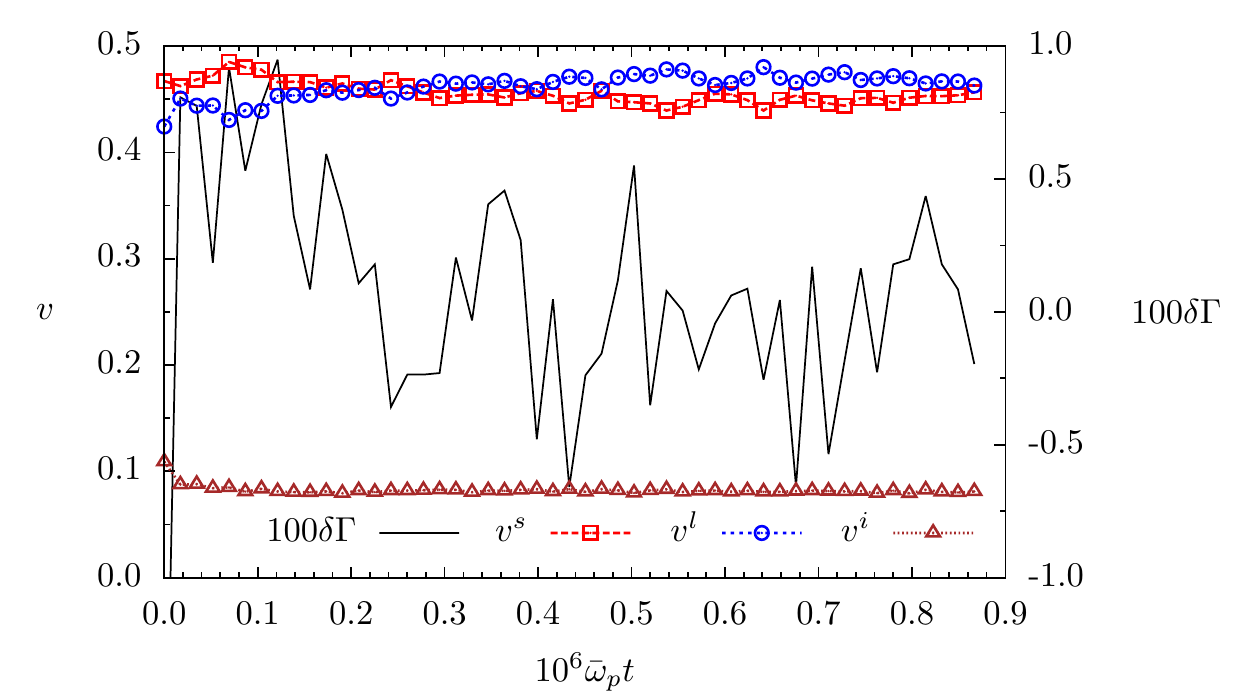}
\caption{(Color on line) Volume fractions of solid (squares), liquid (circles), and interface (triangles) versus time for a 55296 ion simulation that is overall 20\% oxygen and 80\% selenium.  Also shown are fluctuations $\delta\Gamma$ in the Coulomb parameter (solid line), see Eq. \ref{deltaGamma}.}
\label{Fig21}
\end{center}
\end{figure}

\begin{figure}[ht]
\begin{center}
\includegraphics[width=3.5in,angle=0,clip=true] {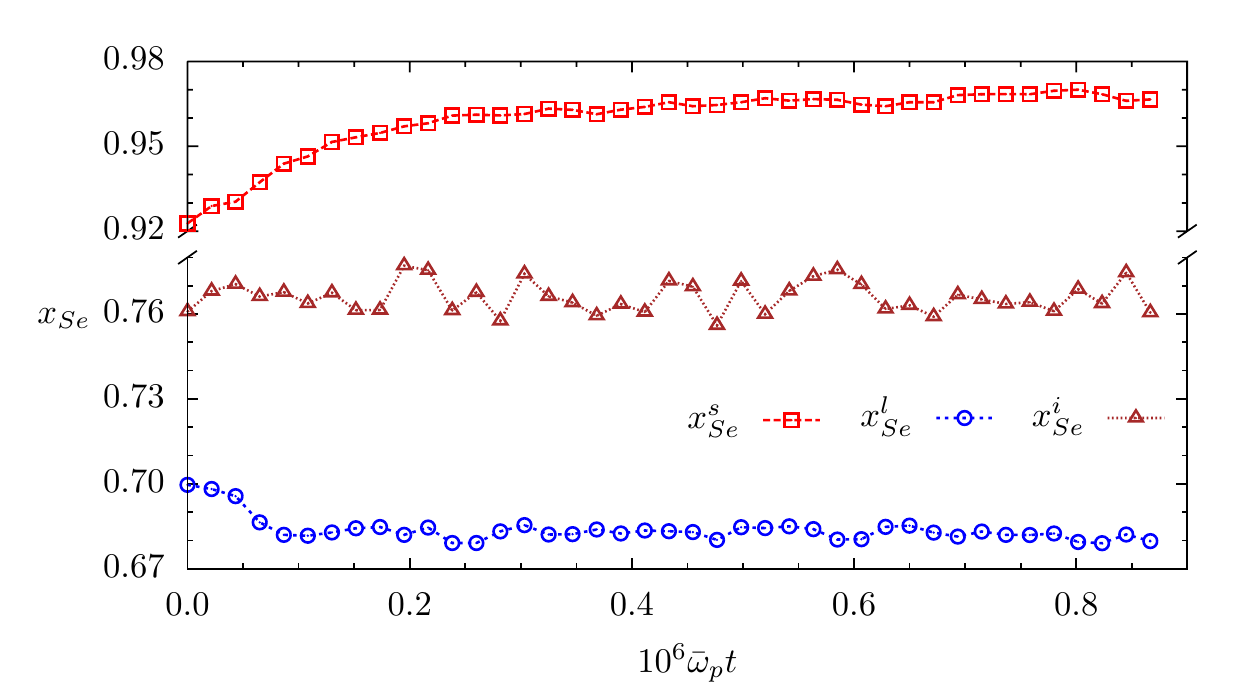}
\caption{(Color on line) Number fraction of selenium $x_{Se}$ in the solid (squares), liquid (circles), and interface (triangles) versus time for a 55296 ion simulation that is overall 20\% oxygen and 80\% selenium.}
\label{Fig22}
\end{center}
\end{figure}

The temperature was adjusted during the run to keep approximately equal volume fractions of solid and liquid as shown in Fig. \ref{Fig21}.   The number fraction of selenium $x_{Se}$ in the solid, liquid and interface are shown in Fig. \ref{Fig22} versus simulation time.  These fractions do not change much in the second half of the run suggesting that the system is (at least approximately) equilibrated.

\begin{figure}[ht]
\begin{center}
\includegraphics[width=3.5in,angle=0,clip=true] {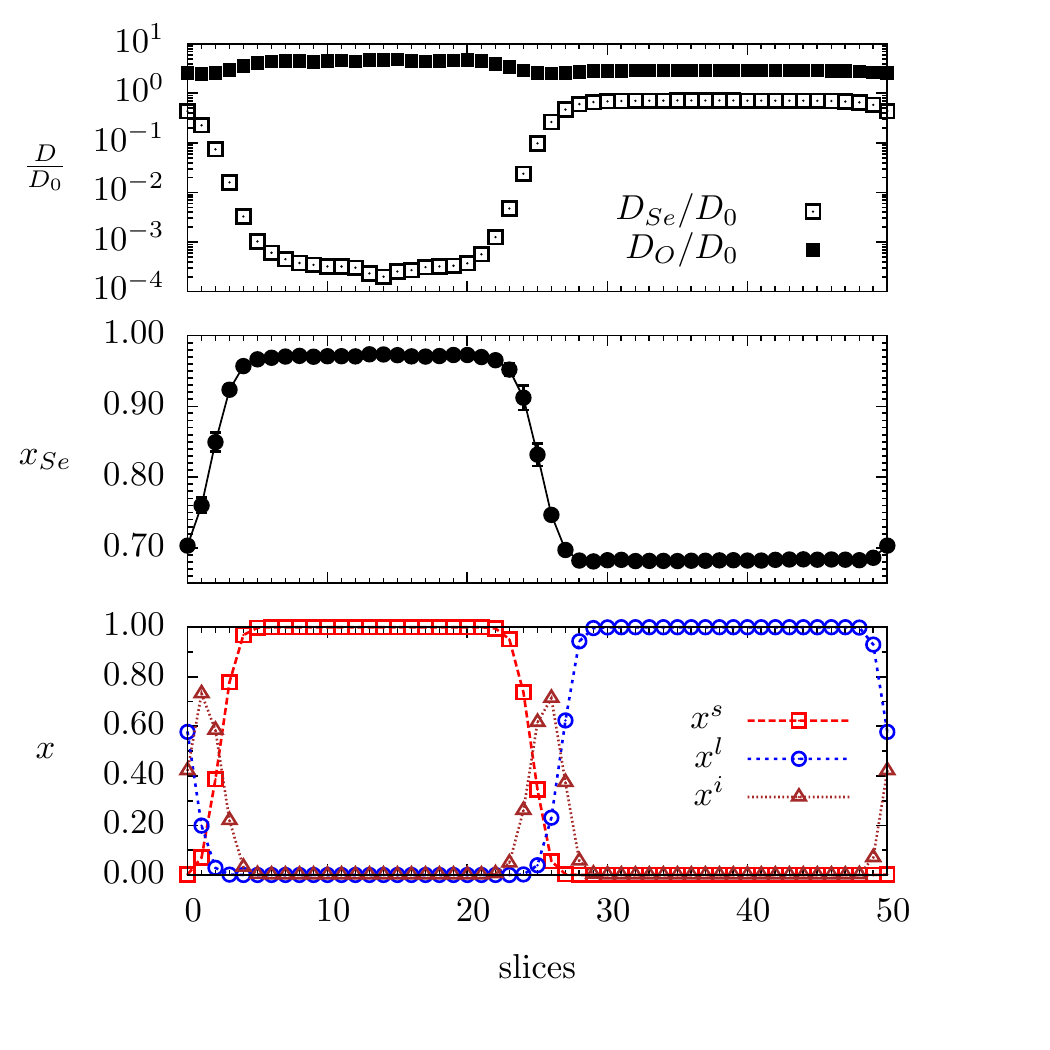}
\caption{(Color on line) Diffusion constant $D$ (top panel)) for selenium (open squares) and oxygen ions (filled squares) for different slices (regions) of the simulation volume. Number fraction of selenium $x_{Se}$ (middle panel) for different slices.  Fraction of ions $x$ (lower panel) that are in the solid (squares), liquid (circles) and interface (triangles)for different slices.  This is an average over the last $1.4\times 10^5\bar\omega_pt$ of a 55296 ion simulation that is overall 20\% oxygen and 80\% selenium.}
\label{Fig23}
\end{center}
\end{figure}

In Fig. \ref{Fig23}, in the middle panel, we show the number fraction of selenium $x_{Se}$ for different slices (regions) of the simulation volume.  This is at the end of the simulation.  In the solid, $x_{Se}$ is seen to be nearly constant and independent of position.  This is consistent with the system having reached equilibration.   Diffusion constants $D$ are also shown in Fig. \ref{Fig23} in the top panel and are seen to behave in a similar way to Fig. \ref{Fig17}. 

\subsection{Run with 70\% Selenium}
\label{subsec.Se70}
The initial conditions for this simulation were prepared similarly to the initial conditions for the 80\% selenium system.  The initial 432 ion solid was, again, composed of 90\% selenium while the 3456 ion liquid was set to 50\% selenium.

\begin{figure}[ht]
\begin{center}
\includegraphics[width=3.5in,angle=0,clip=true] {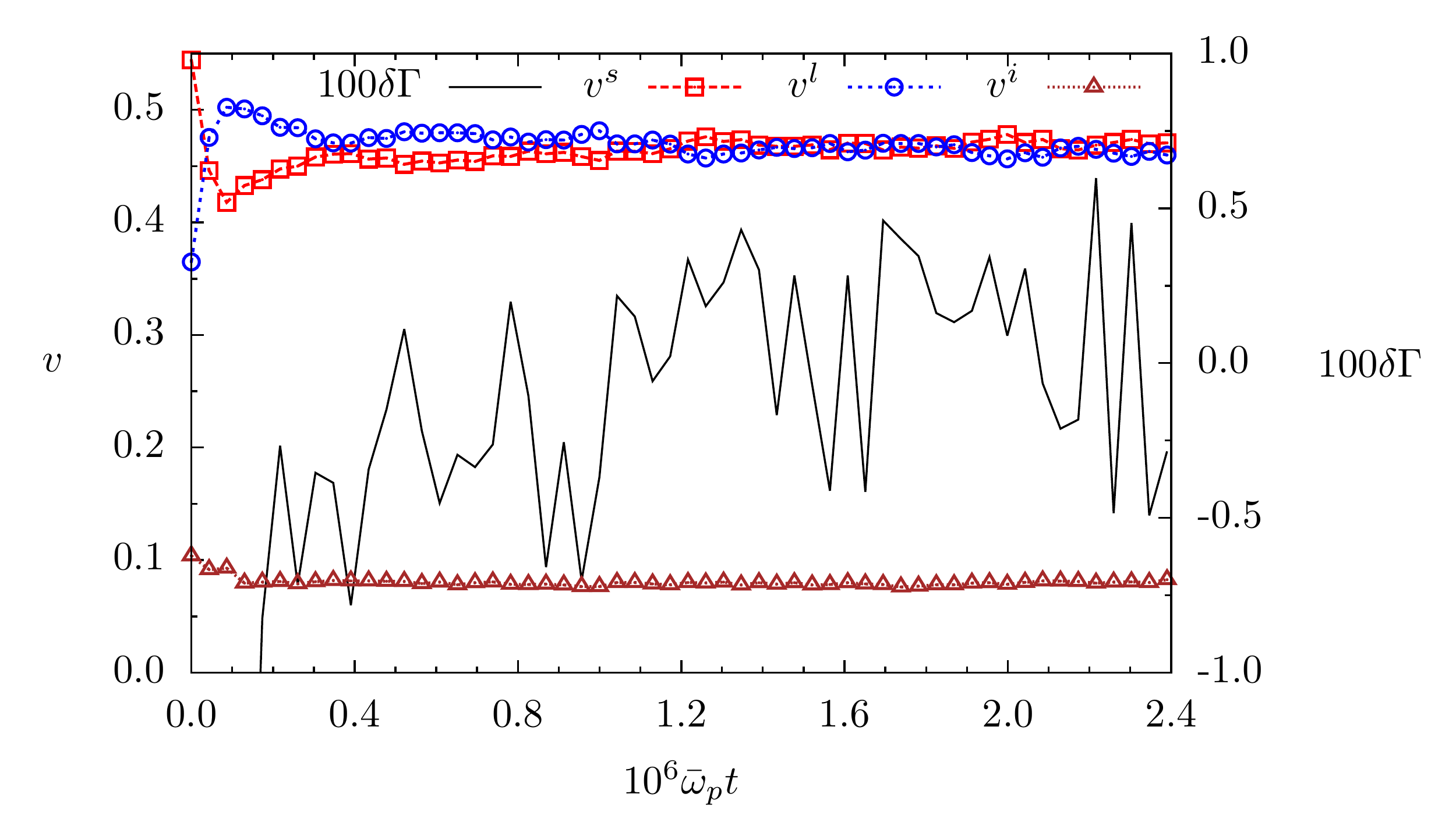}
\caption{(Color on line) Volume fractions of solid (squares), liquid (circles), and interface (triangles) versus time for a 55296 ion simulation that is overall 30\% oxygen and 70\% selenium.  Also shown are fluctuations $\delta\Gamma$ in the Coulomb parameter (solid line), see Eq. \ref{deltaGamma}.}
\label{Fig24}
\end{center}
\end{figure}

\begin{figure}[ht]
\begin{center}
\includegraphics[width=3.5in,angle=0,clip=true] {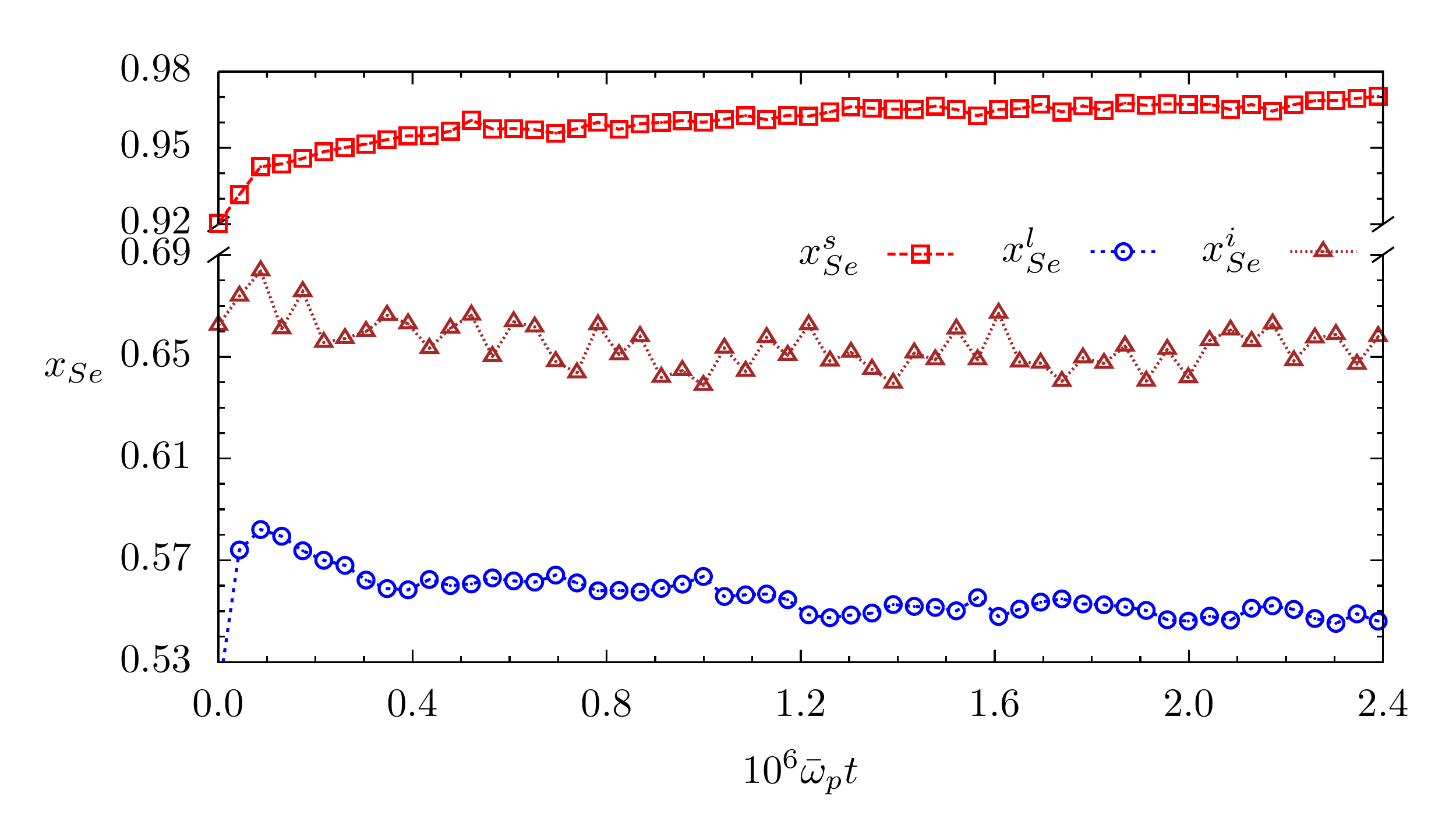}
\caption{(Color on line) Number fraction of selenium $x_{Se}$ in the solid (squares), liquid (circles), and interface (triangles) versus time for a 55296 ion simulation that is overall 30\% oxygen and 70\% selenium.}
\label{Fig25}
\end{center}
\end{figure}

The temperature was adjusted during the run to keep approximately equal volume fractions of solid and liquid as shown in  Fig. \ref{Fig24}.   The number fraction of selenium $x_{Se}$ in the solid, liquid and interface are shown in Fig. \ref{Fig25} versus simulation time.  The fraction of selenium in the solid may be increasing very slowly with time while $x_{Se}$ in the liquid may be slowly decreasing.  Because this change is very slow the system may be near equilibration.  However, it is possible that $x_{Se}$ could continue to change for much longer run times.

\begin{figure}[ht]
\begin{center}
\includegraphics[width=3.5in,angle=0,clip=true] {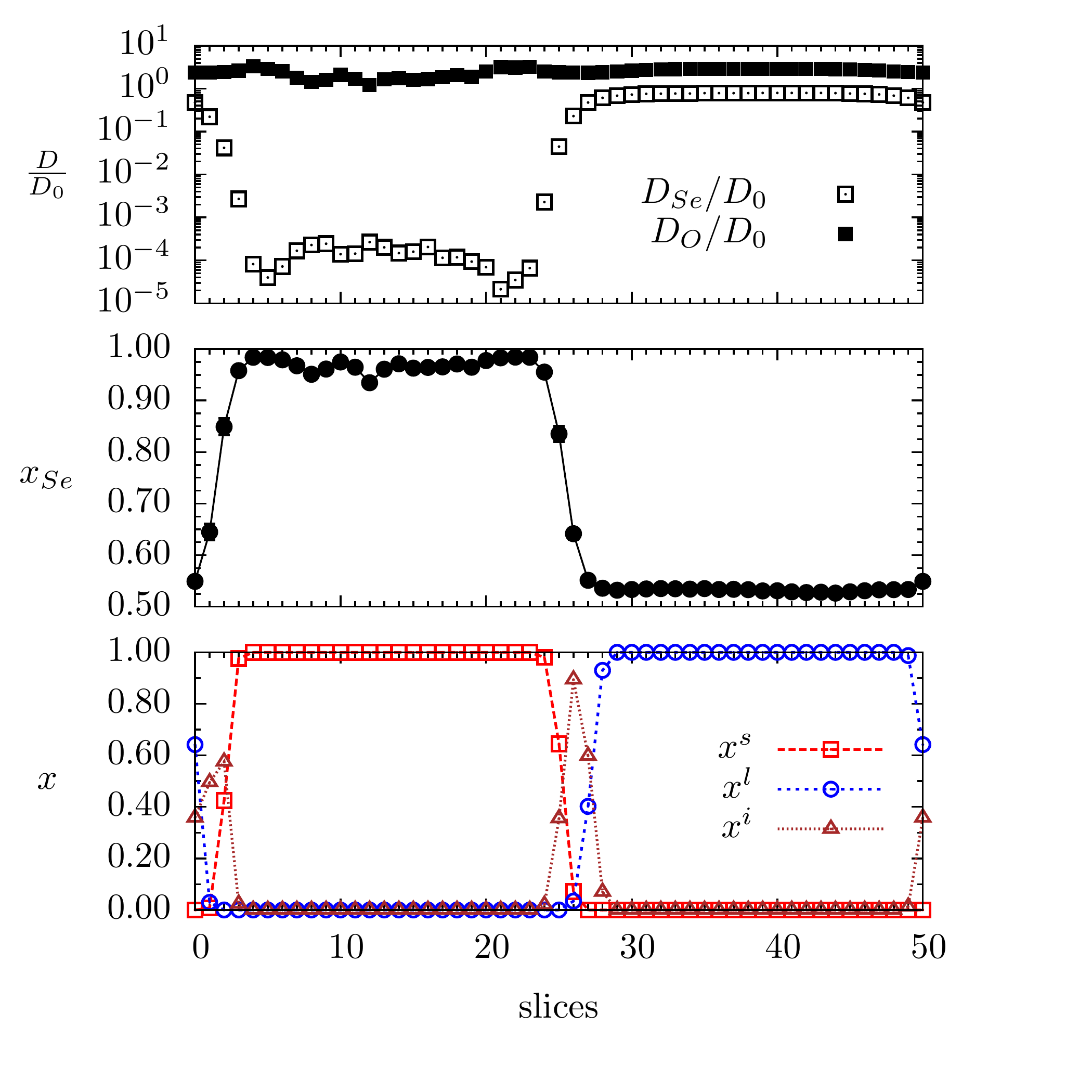}
\caption{(Color on line) Diffusion constant $D$ (top panel)) for selenium (open squares) and oxygen ions (filled squares) for different slices (regions) of the simulation volume. Number fraction of selenium $x_{Se}$ (middle panel) for different slices.  Fraction of ions $x$ (lower panel) that are in the solid (squares), liquid (circles) and interface (triangles)for different slices.  This is an average over the last $1.4\times 10^5\bar\omega_pt$ of a 55296 ion simulation that is overall 30\% oxygen and 70\% selenium.}
\label{Fig26}
\end{center}
\end{figure}

In Fig. \ref{Fig26}, in the middle panel, we show the number fraction of selenium $x_{Se}$ for different slices (regions) of the simulation volume.  This is at the end of the simulation.  In the solid, $x_{Se}$ is seen to be nearly constant and independent of position.  This is consistent with the system having reached equilibration.   Diffusion constants $D$ are also shown in Fig. \ref{Fig26} in the top panel and are seen to behave in a similar way to Fig. \ref{Fig17}.   However now $D$ for selenium is seen to change slightly with position in the solid near the interface regions.

\subsection{Run with 60\% Selenium}
\label{subsec.Se60}
The initial conditions for this simulation were prepared similarly to the initial conditions for the 80\% selenium system.  The initial 432 ion solid was composed of 90\% selenium while the 3456 ion liquid was set to 30\% selenium.

\begin{figure}[ht]
\begin{center}
\includegraphics[width=3.5in,angle=0,clip=true] {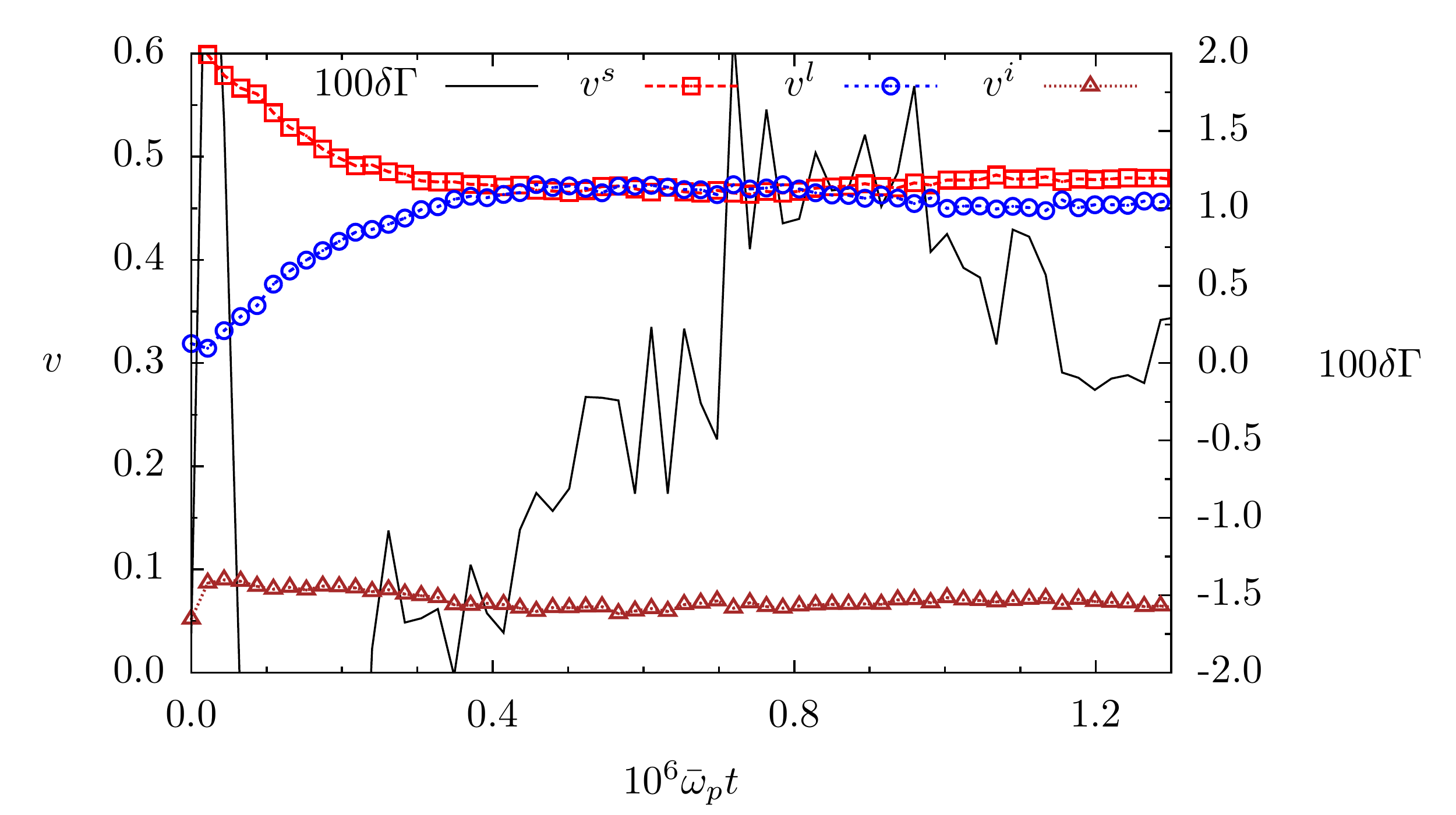}
\caption{(Color on line) Volume fractions of solid (squares), liquid (circles), and interface (triangles) versus time for a 55296 ion simulation that is overall 40\% oxygen and 60\% selenium.  Also shown are fluctuations $\delta\Gamma$ in the Coulomb parameter (solid line), see Eq. \ref{deltaGamma}.}
\label{Fig27}
\end{center}
\end{figure}

\begin{figure}[ht]
\begin{center}
\includegraphics[width=3.5in,angle=0,clip=true] {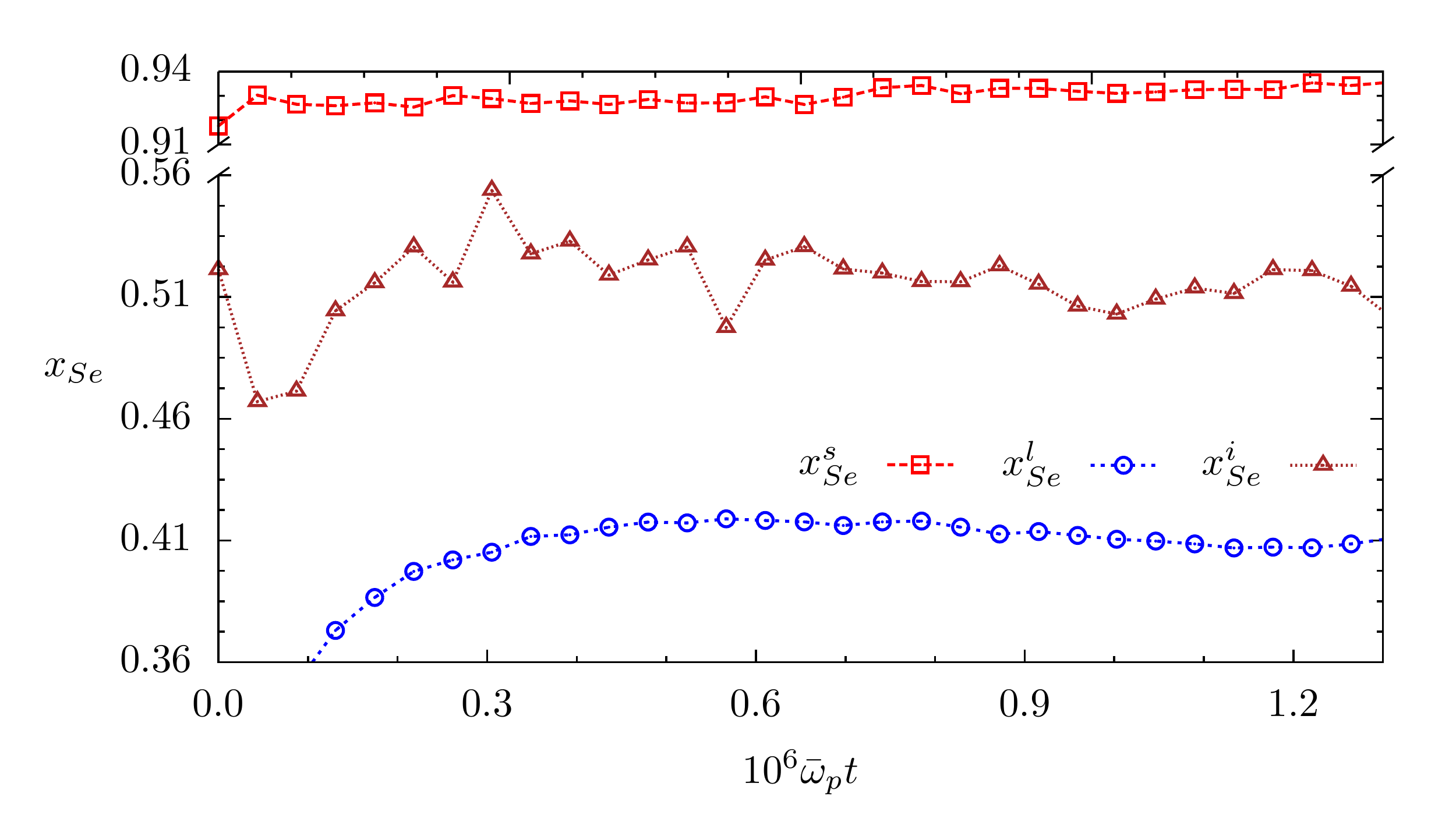}
\caption{(Color on line) Number fraction of selenium $x_{Se}$ in the solid (squares), liquid (circles), and interface (triangles) versus time for a 55296 ion simulation that is overall 40\% oxygen and 60\% selenium.}
\label{Fig28}
\end{center}
\end{figure}

The system started with a large solid fraction.  The temperature was increased to bring the solid fraction approximately equal with the liquid fraction as shown in  Fig. \ref{Fig27}.   The number fraction of senium $x_{Se}$ in the solid, liquid and interface are shown in Fig. \ref{Fig28} versus simulation time.  Again the fraction of Selenium in the solid may be increasing slightly over the second half of the run.  

\begin{figure}[ht]
\begin{center}
\includegraphics[width=3.5in,angle=0,clip=true] {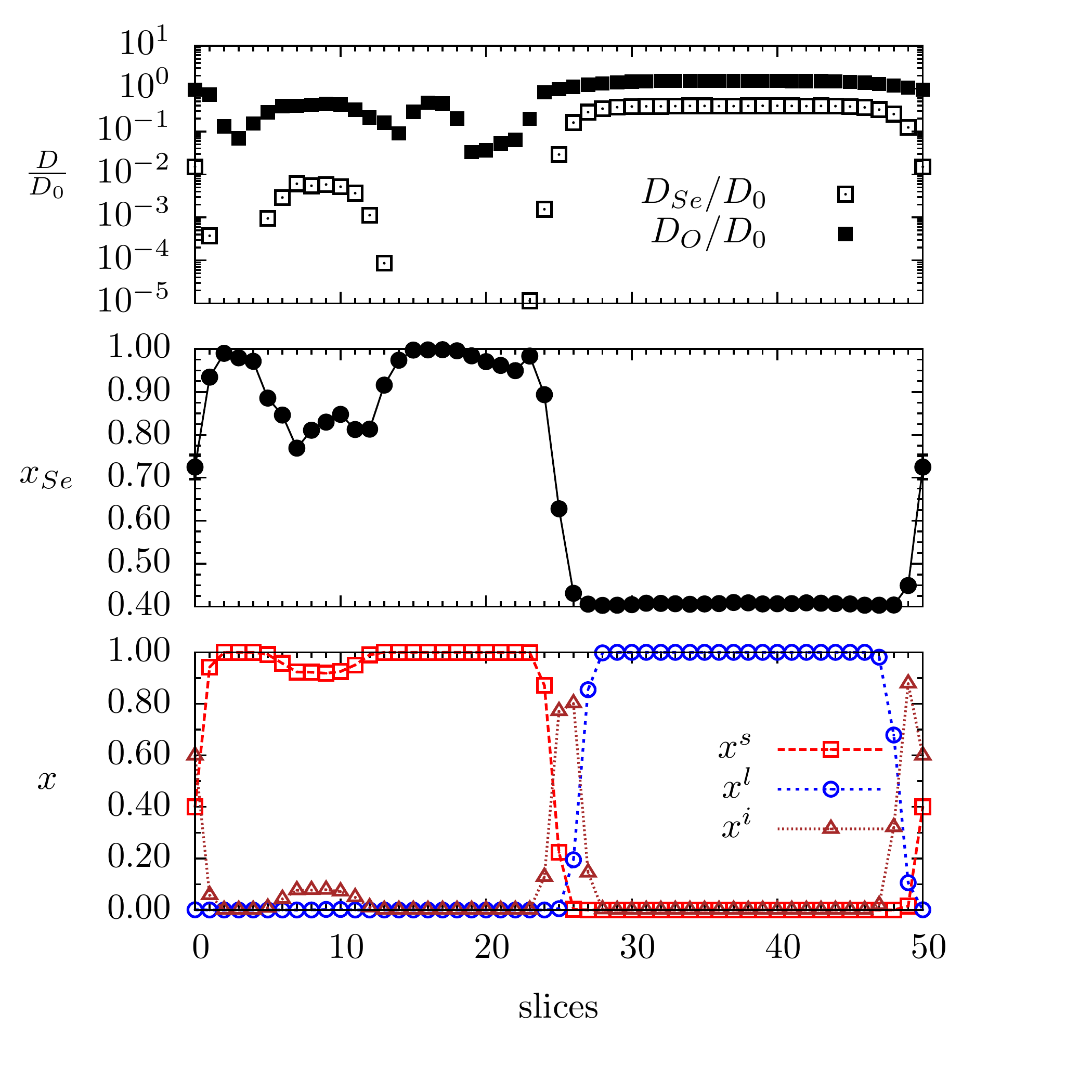}
\caption{(Color on line) Diffusion constant $D$ (top panel)) for selenium (open squares) and oxygen ions (filled squares) for different slices (regions) of the simulation volume. Number fraction of selenium $x_{Se}$ (middle panel) for different slices.  Fraction of ions $x$ (lower panel) that are in the solid (squares), liquid (circles) and interface (triangles)for different slices.  This is an average over the last $1.4\times 10^5\bar\omega_pt$ of a 55296 ion simulation that is overall 40\% oxygen and 60\% selenium.}
\label{Fig29}
\end{center}
\end{figure}

In Fig. \ref{Fig29}, in the middle panel, we show the number fraction of selenium $x_{Se}$ for different slices (regions) of the simulation volume.  This is at the end of the simulation.  In the solid, $x_{Se}$ depends on position, as do diffusion constants $D$  as shown in the top panel of Fig. \ref{Fig29}.  This clearly indicates that the system has not reached equilibrium.  Therefore we can not use this run to determine the phase diagram in Sec. \ref{subsec.OSephasediagram}.  Unfortunately, the equilibration time may be very long for this composition and therefore require an unreasonable amount of simulation time in order to bring the system in equilibration.

\subsection{Run with 50\% Selenium}
\label{subsec.Se50}
The initial conditions for this simulation were prepared similarly to the initial conditions for the 80\% selenium system. The initial 432 ion solid was, again, composed of 90\% selenium while the 3456 ion liquid was set to 10\% selenium.

\begin{figure}[ht]
\begin{center}
\includegraphics[width=3.5in,angle=0,clip=true] {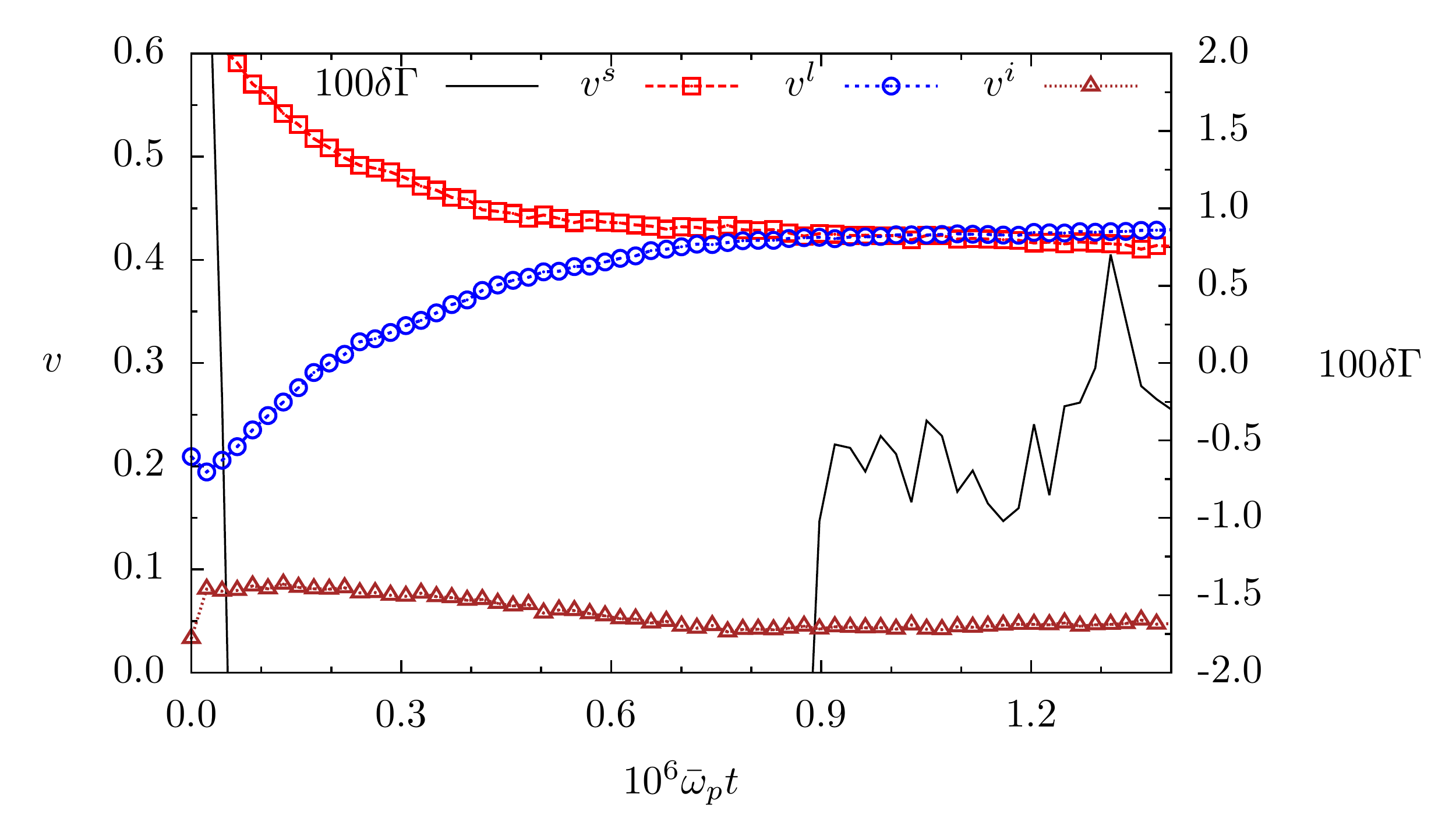}
\caption{(Color on line) Volume fractions of solid (squares), liquid (circles), and interface (triangles) versus time for a 55296 ion simulation that is overall 50\% oxygen and 50\% selenium.  Also shown are fluctuations $\delta\Gamma$ in the Coulomb parameter (solid line), see Eq. \ref{deltaGamma}.}
\label{Fig30}
\end{center}
\end{figure}

\begin{figure}[ht]
\begin{center}
\includegraphics[width=3.5in,angle=0,clip=true] {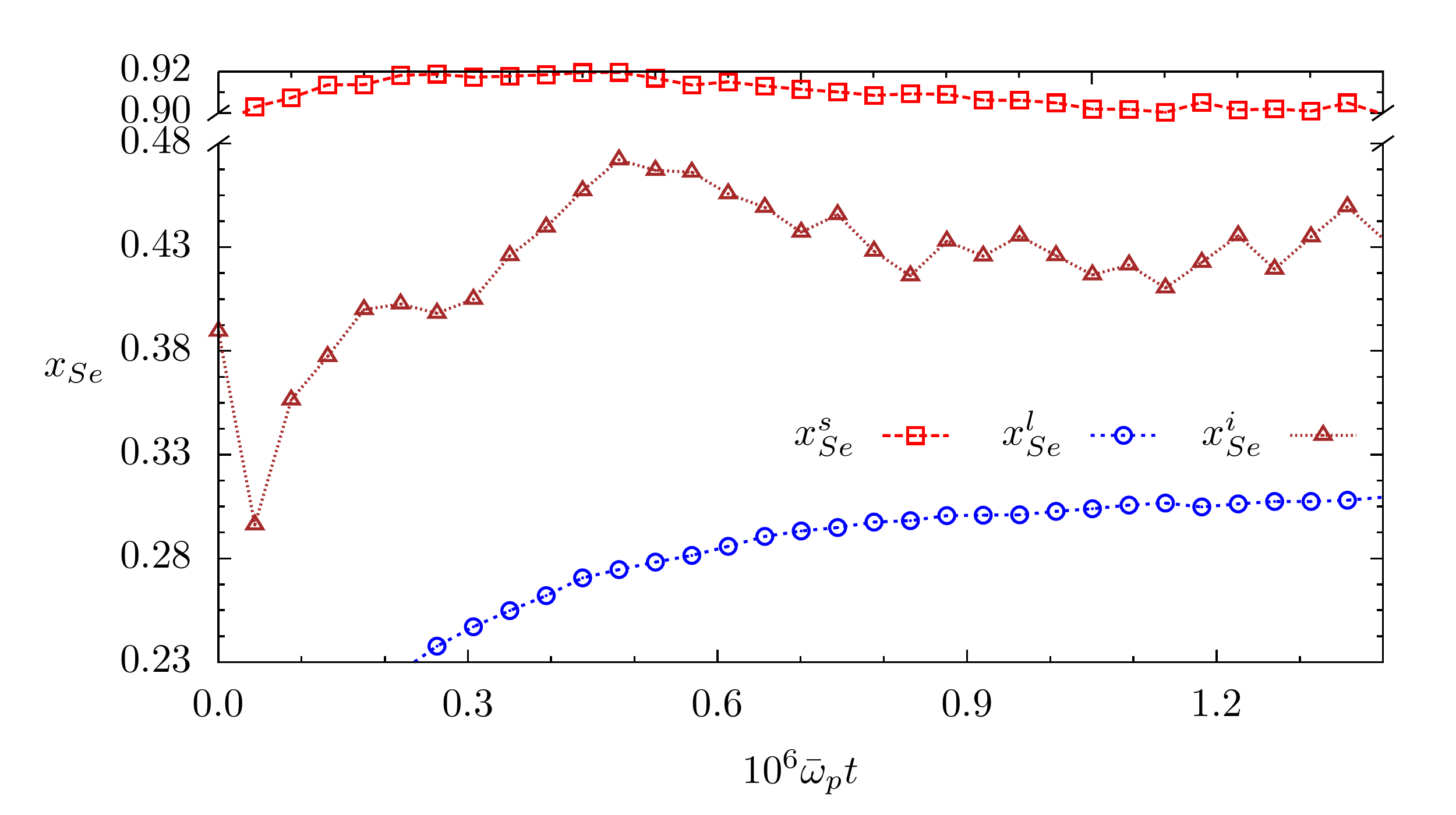}
\caption{(Color on line) Number fraction of selenium $x_{Se}$ in the solid (squares), liquid (circles), and interface (triangles) versus time for a 55296 ion simulation that is overall 50\% oxygen and 50\% selenium.}
\label{Fig31}
\end{center}
\end{figure}

The system started with a large solid fraction.  The temperature was increased to bring the solid fraction approximately equal with the liquid fraction as shown in  Fig. \ref{Fig30}.   The number fraction of selenium $x_{Se}$ in the solid, liquid and interface are shown in Fig. \ref{Fig31} versus simulation time.  Now the fraction of Selenium in the solid may be decreasing slightly, and $x_{Se}$ for the liquid increasing slightly, with time over the second half of the run.  

\begin{figure}[ht]
\begin{center}
\includegraphics[width=3.5in,angle=0,clip=true] {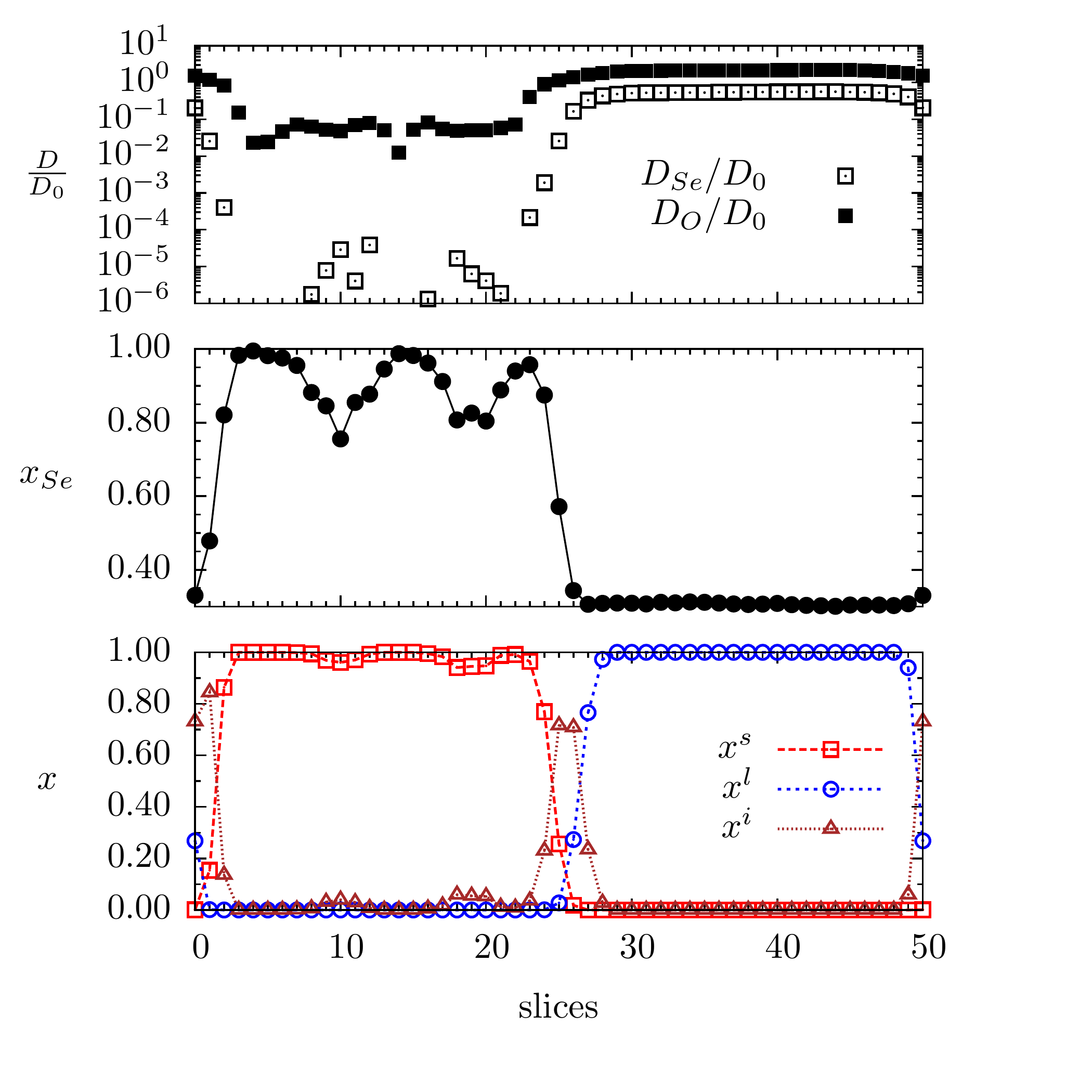}
\caption{(Color on line) Diffusion constant $D$ (top panel)) for selenium (open squares) and oxygen ions (filled squares) for different slices (regions) of the simulation volume. Number fraction of selenium $x_{Se}$ (middle panel) for different slices.  Fraction of ions $x$ (lower panel) that are in the solid (squares), liquid (circles) and interface (triangles)for different slices.  This is an average over the last $1.4\times 10^5\bar\omega_pt$ of a 55296 ion simulation that is overall 50\% oxygen and 50\% selenium.}
\label{Fig32}
\end{center}
\end{figure}

In Fig. \ref{Fig32}, in the middle panel, we show the number fraction of selenium $x_{Se}$ for different slices (regions) of the simulation volume.  This is at the end of the simulation.  In the solid, $x_{Se}$ depends on position, as do diffusion constants $D$  as shown in the top panel of Fig. \ref{Fig32}.  This is similar to the run with 60\% selenium and clearly indicates that the system has not reached equilibrium.  Therefore we can not use this run to determine the phase diagram in Sec. \ref{subsec.OSephasediagram}.  Unfortunately, the equilibration time for this composition may also be very long and therefore require an unreasonable amount of simulation time in order to bring the system in equilibration.

\subsection{Oxygen-Selenium Phase diagram}
\label{subsec.OSephasediagram}
\begin{figure}[ht]
\begin{center}
\includegraphics[width=3.5in,angle=0,clip=true] {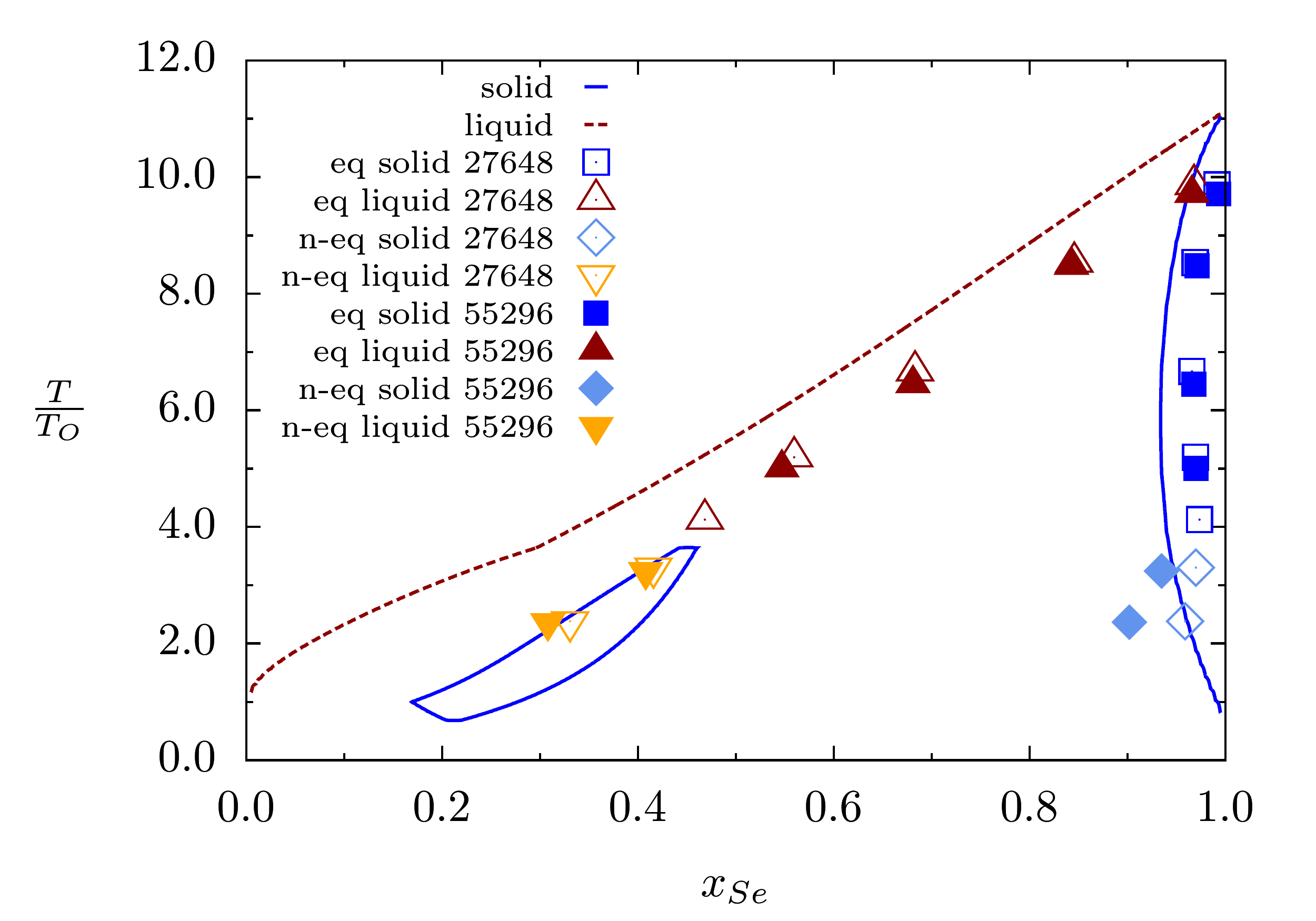}
\caption{Selenium-oxygen phase diagram plotting the composition of the liquid phase (upper red curve or triangles) that is in equilibrium with the solid phase (lower blue curve or squares).  Results from 55296 ion simulations are filled symbols while the open symbols are results with 27648 ions.  Besides results discussed in the text there is an extra 27648 simulation with $x_{Se} = 0.65$.  The curves are the results of Medin and Cumming \cite{medin}.}
\label{Fig33}
\end{center}
\end{figure}

We now present the liquid-solid phase diagram implied by the simulations in Sections \ref{subsec.Se98}, \ref{subsec.Se90}, \ref{subsec.Se80}, \ref{subsec.Se70}, \ref{subsec.Se60}, and \ref{subsec.Se50}.  We use the data in Table \ref{tablethree}.  Figure \ref{Fig33} shows the phase diagram as a function of $x_{Se}$.  The $y$ axis is the melting temperature $T$ divided by the melting temperature $T_O$ for pure oxygen.  We assume the pure oxygen system melts at $\Gamma_m=178.4$ \cite{WDPRL}.  This differs slightly from the one component plasma result because we include the effects of electron screening.

The filled upward pointing triangles in Fig. \ref{Fig33} show the composition of the liquid phase, and the filled squares the composition of the solid phase for 55296 ion simulations.  Also shown as filled downward pointing triangles, and filled diamonds are the liquid and solid compositions for runs that are clearly not equilibrated.  These points should not be used in the determination of the phase diagram.  The open triangles (liquid) and squares (solid) show results for smaller 27648 ion simulations, see Table \ref{tablefive}.  In general there is very good agreement between equilibrated 55296 and 27648 ion simulations.  This suggests that finite size effects, while not strictly zero, are small.

 \begin{table}[ht]
\begin{tabular}{|c||c|ccc|}
\hline
$x_{Se}$&$\bar{\Gamma}$&$x_{Se}^s$&$x_{Se}^l$&$x_{Se}^i$\\
\hline
0.98 &198.1(7)&0.991(1)&0.968(1)&0.976(2)\\
0.90 &212.0(8)&0.969(2)&0.846(4)&0.888(4)\\
0.80 &244(2)&0.966(2) &0.683(3)&0.764(6)\\
0.70 &279(1)  &0.970(2) &0.560(3)&0.650(8)\\
0.65 &329(1)  &0.973(2) &0.469(3)&0.559(6)\\
0.60 &383(1)  &0.971(2) &0.415(3)&0.487(7)\\ 
0.50 &456(1)  &0.959(1) &0.330(1)&0.400(6)\\
\hline
\end{tabular}
\caption{Equilibrium compositions of 27648 ion runs. 
Selenium number fraction of the whole system is $x_{Se}$.  Note that runs with $x_{Se}=0.60$ and 0.50 are not equilibrated.  The Coulomb parameter averaged over the last third of the run is $\bar{\Gamma}$. 
The composition of the solid is $x_{Se}^s$, the liquid is $x_{Se}^l$, and the interface regions is $x_{Se}^i$.
Statistical errors are quoted in the last digit in parentheses.}
\label{tablefive}
\end{table}

The blue solid lines in Fig. \ref{Fig33} show the solid composition, and the dotted brown line the liquid composition, for the phase diagram of Medin and Cumming \cite{medin}.  There is qualitative agreement between these curves and our results for the $x_{Se}>0.5$ half of the phase diagram where we have apparently equilibrated independent results.  However there are some differences in detail.  We find a somewhat larger selenium solid composition while our melting temperature for the liquid is somewhat lower than Medin and Cumming.   This difference in melting temperature may be due to electron screening effects that are included in our simulations and neglected in Medin and Cumming's free energies.  

\begin{figure}[ht]
\begin{center}
\includegraphics[width=3.5in,angle=0,clip=true] {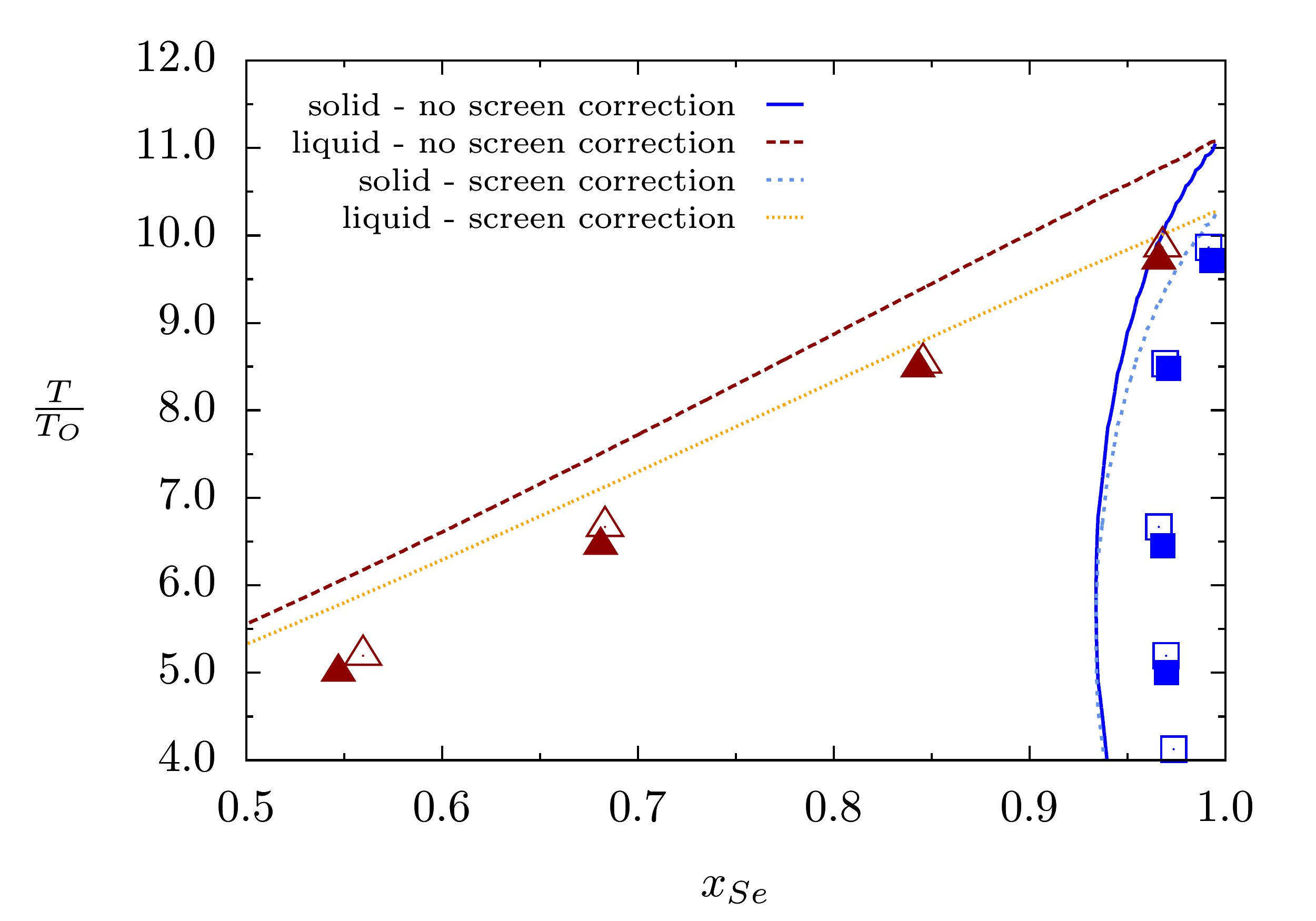}
\caption{Enlargement of the large $x_{Se}$ part of the Selenium-oxygen phase diagram from Fig. \ref{Fig33}. The heavy dashed and solid lines are the results of Medin and Cumming \cite{medin}, while the light dashed lines are our rescaling of Medin and Cumming's results to approximately include the effects of electron screening as discussed in the text.}
\label{Fig34}
\end{center}
\end{figure}

Screening depends on the ratio of ion sphere radius $a$ to screening length $\lambda$, see Eq. \ref{v(r)}.  For a very relativistic electron gas this ratio depends only on average ion charge $\langle Z \rangle$ and is independent of density,
\begin{equation}
\kappa = \frac{a}{\lambda}=\bigl(\frac{2^\frac{1}{3}3^\frac{2}{3}}{\pi^\frac{1}{6}}\bigr)\alpha^\frac{1}{2}\langle Z \rangle^\frac{1}{3}\, .
\end{equation} 
For a one component Yukawa system the value of $\Gamma$ at the melting point, $\Gamma_m$, is \cite{hamaguchi}
\begin{equation}
\Gamma_m(\langle Z \rangle) \approx 171.8 + 42.46 \kappa^2 + 3.841 \kappa^4,
\label{Gamma_m}
\end{equation}
for $\kappa\leq 1.4$.  For pure oxygen we have
\begin{equation}
\Gamma_m(8)=178,
\label{Gamma_O}
\end{equation}
while for pure selenium,
\begin{equation}
\Gamma_m(34)=188.
\label{Gamma_Se}
\end{equation}
Thus pure selenium melts at a 6\% higher $\Gamma$ value than pure oxygen because of enhanced electron screening.  In order to study the differences between our MD simulation results for $T/T_O$ in Fig. \ref{Fig33} and Medin and Cumming's results, we rescale Medin and Cumming's melting temperatures according to
\begin{equation}
\frac{T}{T_O} \rightarrow \frac{\Gamma_m(8)}{\Gamma_m(\langle Z \rangle)}\, \frac{T}{T_O}.
\label{rescale}
\end{equation}
 and plot the rescaled results in Fig. \ref{Fig34}.  This procedure ensures that the rescaled Medin and Cumming results will reproduce Eq. \ref{Gamma_O} for pure oxygen and Eq. \ref{Gamma_Se} for pure selenium.  In between, for a mixture of oxygen and selenium, we somewhat arbitrarily assume that the electron screening correction depends only on $\langle Z \rangle$.

There is good agreement between our MD results in Fig. \ref{Fig34} and these rescaled Medin and Cumming results for simulations with 90\% or 98\% selenium.  However for simulations with smaller $x_{Se}$, our MD results give somewhat lower melting temperatures and have less oxygen in the solid phase.  These differences could be due to finite size or non-equilibrium effects in our MD simulations.  However, we only find very small differences between 27648 and 55296 ion simulations.   Furthermore, except for the low $x_{Se}=50$\% or 60\%  runs, we find very little time dependence in our simulations.  Alternatively the differences could be due to electron screening effects for mixtures that are not well described by the rescaling in Eq. \ref{rescale}.  Finally, the differences could be due to limitations in the liquid and solid free energies used by Medin and Cumming.  

Unfortunately our simulations with 50\% and 60\% selenium did not reach equilibrium.  Therefore we are not able to effectively study the phase diagram for low selenium concentrations.  Medin and Cumming predict regions of the phase diagram with equilibria between different solid phases.  Perhaps by carefully preparing initial conditions that include two solid phases of different compositions, one may be able to study solid-solid phase equilibria with our direct MD simulation procedure.  However, the small diffusion constants for selenium in the solid have made it difficult for us to equilibrate the simulations with small selenium concentrations presented here.  Note that this region of the phase diagram, with small $x_{Se}$, may not be important in applications for neutron stars. 

The complex rapid proton capture nucleosynthesis ash composition considered in ref. \cite{HBB} was predominately selenium, with only small concentrations of oxygen and a number of other impurities.  We modeled this 17 component composition with the binary system of oxygen impurities mixed with the dominant element selenium.  Direct MD simulations of the full rp ash composition in ref. \cite{HBB} Table I found the concentration of oxygen in the liquid phase to be six times larger than the oxygen concentration of the solid phase.  While we find, in the first two rows of Table \ref{tablethree}, that the oxygen concentration of the liquid phase, for our simplified binary mixture simulations, to be five times larger than that in the solid phase.  We conclude that this binary mixture model provides a reasonable description of the freezing behavior of the rp ash.

\section{Summary and Conclusions}
\label{Conclusions}
We have determined the liquid-solid phase diagram for carbon-oxygen and oxygen-selenium plasma mixtures using two-phase MD simulations.  We identified liquid, solid, and interface regions in our simulations using a bond angle metric described in Sec. \ref{subsec.phase}.  To study finite size effects, we performed both 27648 and 55296 ion simulations.  To help monitor non-equilibrium effects, we calculated diffusion constants $D_i$.  For the carbon-oxygen system, we find that $D^s_O$ for oxygen ions in the solid is much smaller than $D^s_C$ for carbon ions and that both diffusion constants are 80 or more times smaller than diffusion constants in the liquid phase.  There is excellent agreement between our carbon-oxygen phase diagram and that predicted by Medin and Cumming \cite{medin}.  This suggests that errors from finite size and non-equilibrium effects are small, and that the carbon-oxygen phase diagram is now accurately known.    

The oxygen-selenium system, with a larger ratio of charges than carbon-oxygen, can serve as a simple two component model of the complex rapid proton capture ash composition on an accreting neutron star.  We find that diffusion of oxygen in a predominately selenium crystal is remarkably fast and is comparable to diffusion in the liquid phase.  Our MD simulations have a somewhat lower melting temperature for the oxygen-selenium system than that predicted by Medin and Cumming.  This is in part due to electron screening effects, that are included in our simulations and may be neglected by Medin and Cumming.  In the future, we will present MD simulations of the phase diagram for the three component carbon-oxygen-neon system to include the effects of neon impurities in carbon-oxygen white dwarfs.

We thank Z. Medin for helpful discussions.  This research was supported in part by DOE grant DE-FG02-87ER40365 and by the National Science Foundation through TeraGrid resources provided by the National Institute for Computational Sciences under grant TG-AST100014.

\vfill\eject

\end{document}